\documentclass[fleqn,usenatbib]{mnras}

\usepackage{newtxtext,newtxmath}
\usepackage{threeparttable}
\usepackage{hyperref}
\usepackage{soul}
\hypersetup{colorlinks,linkcolor={blue},citecolor={blue},urlcolor={blue}}

\usepackage[T1]{fontenc}

\DeclareRobustCommand{\VAN}[3]{#2}
\let\VANthebibliography\thebibliography
\def\thebibliography{\DeclareRobustCommand{\VAN}[3]{##3}\VANthebibliography}

\usepackage{graphicx}	
\usepackage{amsmath}	

\graphicspath{{./}{plots/}}

\title[Multi-messenger emission across jet scales]{The role of dissipation distance on reconnection-driven multi-messenger signals from blazar jets}

\author[Stathopoulos \& Petropoulou]{
S. I. Stathopoulos,$^{1,2}$\thanks{E-mail: stamstath@phys.uoa.gr}
Maria Petropoulou$^{1,2}$
\\
$^{1}$Department of Physics, National and Kapodistrian University of Athens, University Campus Zografos, GR 15784, Athens, Greece\\
$^{2}$Institute of Accelerating Systems \& Applications, University Campus Zografos, Athens, Greece\\ 
}

\date{Accepted XXX. Received YYY; in original form ZZZ}

\pubyear{\the\year{}}

\begin{document}
\label{firstpage}
\pagerange{\pageref{firstpage}--\pageref{lastpage}}
\maketitle

\begin{abstract}

Blazars are characterized by relativistic jets that are closely aligned with our line of sight. This results in relativistic beaming, making blazars among the most luminous extragalactic sources across the electromagnetic spectrum, from radio waves to $\gamma$-rays and, potentially, in high-energy neutrinos. We present a comprehensive study of multi-messenger emission from blazar jets powered by magnetic reconnection occurring at varying distances from the supermassive black hole (SMBH). By generalizing previous models, we explore how the emission characteristics depend self-consistently on the spatial evolution of key jet properties, including magnetization, bulk Lorentz factor, and external photon fields (accretion disc, broad-line region, and dusty torus). Using numerical simulations, we examined the impact of the initial jet magnetization, particle acceleration efficiency, jet-to-accretion power ratio, and mass accretion rate on the broadband photon spectra and neutrino emission. Our findings reveal distinct emission regimes characterized by different dominant radiative processes: synchrotron and synchrotron self-Compton dominate closer to the SMBH where magnetization is high, while external Compton (EC) processes become significant near the broad-line region (BLR). Neutrino production efficiency is highest upstream of the BLR, driven by enhanced photon target densities from synchrotron and external photons available for photopion interactions, whereas the proton particle distribution is hard. Our model predictions are compared with observations of $\gamma$-ray luminosities and synchrotron peak energies of Fermi-detected blazars, highlighting magnetic reconnection as a potential mechanism driving both electromagnetic and neutrino emissions in astrophysical jets.

\end{abstract}

\begin{keywords}
Astroparticle physics, Methods: numerical, Radiation mechanisms: non-thermal, Radiative transfer
\end{keywords}



\section{Introduction}
Blazars represent a distinct subclass of active galactic nuclei (AGN), characterized by relativistic jets oriented at small angles to our line of sight \citep{Urry1995}. These objects rank among the most luminous and persistent sources of non-thermal electromagnetic emission in the Universe, exhibiting a broadband spectrum with a characteristic two-hump morphology \citep[for a review, see][]{Padovani2017}. Blazars are the most numerous class of persistent extragalactic sources in the gamma-ray band \citep{Abdollahi2020}, and have been proposed as potential sites for cosmic-ray acceleration and production of high-energy neutrinos \citep[e.g.][]{Stecker1991, Mannheim1995, Muecke2003, Atoyan2004}. Notably, the first astrophysical source associated with high-energy neutrinos was blazar TXS~0506+056 \citep{IceCube2018_flare, IceCube2018_archival}.

AGN jets are believed to originate near rotating, accreting black holes, emerging as plasma outflows dominated by Poynting flux \citep{1977MNRAS.179..433B}. In the framework of ideal magnetohydrodynamics (MHD), theory and simulations suggest that the jet's bulk acceleration occurs through the gradual conversion of magnetic energy, quantified by its magnetization $\sigma$ (the ratio of Poynting flux to the total energy flux of the jet). As a result, the bulk Lorentz factor of the jet $\Gamma$ increases while $\sigma$ decreases, maintaining a constant total energy flux in the absence of dissipative processes. This magnetic acceleration is a gradual process that continues until the jet transitions from being magnetically dominated  ($\sigma \gg 1$) to being kinetically dominated  ($\sigma < 1$) \citep[e.g.][]{Vlahakis2004, Beskin2006,Komissarov2007,Komissarov2011}.

If jets were ideal MHD unperturbed outflows without any mechanism for tapping the available jet energy and transferring it to non-thermal particles, they would simply be dark. Therefore, energy dissipation mechanisms should be in place in jets. Several instabilities can perturb the outflow and lead to changes in the topological arrangement of the magnetic field lines \citep[e.g.][]{2008MNRAS.384L..19B, 2017MNRAS.469.4957B,2018PhRvL.121x5101A, Giannios2019, 2021ApJ...912..109K}. The opposite orientation of the magnetic fields can lead to magnetic reconnection, which is a promising mechanism for accelerating charged particles to relativistic energies. This mechanism has gained further support from X-ray polarimetry studies -- especially in high‑synchrotron‑peaked (HSP) blazars, showing strong and variable polarization signatures indicative of ordered magnetic structures and reconnection zones in the jet core~\citep{2021Galax...9...37T}. The distribution function of accelerated particles during relativistic magnetic reconnection can be well described by a power law with an exponential cutoff at high energies \citep[see][for recent reviews]{2024SSRv..220...43G, 2025arXiv250602101S}. Interestingly, simulations of magnetic reconnection show that the power-law slope depends on the magnetization of the plasma \citep[e.g.][]{2014PhRvL.113o5005G, 2014ApJ...783L..21S, Werner2016}, whereas non-thermal particles receive a significant fraction of the dissipated magnetic energy regardless of the exact value of $\sigma$, as long as $\sigma \gg 1$ \citep[e.g.][]{2015MNRAS.450..183S, 2019ApJ...880...37P}. If magnetic reconnection is initiated at varying locations along the jet, where the bulk properties of the flow, such as magnetization and Lorentz factor, are also expected to vary, then the resulting radiative output from each dissipative event is likewise expected to exhibit distinct characteristics. Recent work has taken this idea further by showing that the same jet, if ``activated'' at various distances, can reproduce the full diversity of blazar flares, from orphan optical and $\gamma$-ray outbursts to fully multi-wavelength events, within a single stochastic-dissipation framework \citep{2022PhRvD.105b3005W}.

\cite{Petro2023} investigated this by looking at the photon and neutrino emission expected from blazar jets, if reconnection was triggered close to the BLR, assuming jets with different $\sigma, \Gamma$ values at the same location. They considered electron–proton jets and related the jet’s bulk Lorentz factor ($\Gamma$) and magnetization ($\sigma$) through a constant energy-per-baryon $\mu = \Gamma(1+\sigma)$. An empirical link between $\Gamma$ and the accretion rate (and hence BLR/disk luminosity) was used so that high-accretion jets have intense external photon fields, while low-accretion jets have weak external fields. They found that the blazar spectral energy distribution (SED) was predominantly produced by electrons via synchrotron and inverse Compton (IC) processes, with hadronic emission (and thus neutrino production via photopion interactions) generally subdominant except for high magnetizations in the emitting region. The dissipation region was assumed to be fixed close to the BLR, where abundant external photons serve as targets for IC scattering and photopion interactions, and thus neutrino production. In this framework, low-luminosity blazars (BL Lac–like, $L_{\gamma}\lesssim 10^{45}$ erg s$^{-1}$) were associated to less powerful, slower jets that remain highly magnetized at the dissipation site, resembling a BL Lac–type SEDs with neutrino emission ($L_{\nu}\sim0.3$–$1 L_{\gamma}$). In contrast, high-luminosity blazars (FSRQ-like, $L_{\gamma}\gg10^{45}$ erg s$^{-1}$) were shown to be powered by faster jets with lower $\sigma$ by the time they dissipate. An earlier reconnection-powered blazar model~\citep{2015MNRAS.449..431J} similarly argued that the extreme $\gamma-$ray dominance of luminous FSRQs requires a low-$\sigma$ emission zone, implying that the high-energy emission in blazars may produced in weakly magnetized jet regions.

In this study, we generalize the model introduced by \cite{Petro2023} by relaxing the assumption of a fixed location for the emitting region (near the BLR) to perform a more comprehensive and less biased parametric exploration. Specifically, we expand the emission model by incorporating varying locations for energy dissipation along the jet, self-consistently accounting for changes in jet bulk properties (magnetization and Lorentz factor) and external photon fields (e.g., accretion disc AD, BLR, and dusty torus DT). By exploring different dissipation distances and the corresponding bulk properties of the jet, we aim to a deeper understanding of blazar multi-messenger emissions and to more realistic predictions for the photon and neutrino emission from blazar jets.

This manuscript is structured as follows. In Sec.~\ref{sec:model}, we detail the jet dynamics, particle acceleration mechanisms, and radiative processes used in our model. In Sec.~\ref{sec:methods}, we describe our numerical approach and methods. Results are presented in Sec.~\ref{sec:results} and \ref{sec:neutrino_prod}, where we discuss implications for photons and the neutrino spectrum in our model. Sec.~\ref{sec:observations} compares model predictions with observations of blazars. Finally, we summarize and discuss our results in Sec.~\ref{sec:discussion} and conclude our findings in Sec.~\ref{sec:conclusions}.
 
\section{Model}\label{sec:model}
We present a radiative model for blazar jets that accounts for the dependence of the jet bulk properties with distance from the black hole. We investigate the radiation in photons and neutrinos emitted by plasma in a perturbed region of the jet, where magnetic reconnection occurs. This analysis includes the effects of external photon fields, which vary depending on the location of the emitting region in the jet. Additionally, we account for the jet dynamics, including the evolution of its geometry, bulk motion, and magnetization, as a function of distance from the SMBH. We further examine how the jet bulk properties influence particle acceleration and the resulting observable emissions. By varying the distance of the emitting region, we explore how external photon fields modulate the jet's radiation output and other observable properties, such as neutrino emission. The description of external fields is described in Appendix \ref{app:external}.

\subsection{Jet dynamics}
In this section, we define the global parameters of the (unperturbed) jet as functions of the distance $z$ from the base of the SMBH, including its geometry, bulk flow evolution, and magnetic field strength. 
We assume that the jet acceleration begins at height $z_0$ with an initial Lorentz factor $\Gamma_{0} \gtrsim 1$ (in all calculations that follow we set $\Gamma_0 = 1.1$) and is driven by magnetocentrifugal acceleration, assuming a rotating SMBH \citep{1977MNRAS.179..433B}. The Lorentz factor evolves with distance as,
\begin{equation} 
\Gamma(z) = \Gamma_{0} + (\Gamma_{\max} - \Gamma_{0}) \frac{z^{1/2} - z_0^{1/2}}{z_{\rm acc}^{1/2} - z_0^{1/2}},  z \le z_{\rm acc}
\label{Eq:G_z}
\end{equation} 
and $\Gamma = \Gamma_{\max}$ for larger distances, where $\Gamma_{\max}$ is the maximum Lorentz factor attained at distance $z_{\rm acc}$. The Doppler factor of the moving plasma at distance $z$ is given by
\begin{equation} 
\delta = \frac{1}{\Gamma(z) (1- \beta(z) \cos \theta_{\rm obs})}
\label{Eq:delta}
\end{equation} 
where $\theta_{\rm obs}$ is the angle between the observer's line of sight and the jet axis. The jet cross-section radius $R(z)$ is given by,
\begin{equation} 
R(z) = R_0 + (z - z_0) \tan\left( \frac{\eta_0}{\Gamma(z)} \right), 
\label{Eq:R_z} 
\end{equation}
where $R_0$ is the initial jet radius and $\eta_0$ is the apparent jet's opening angle. The term $\frac{\eta_0}{\Gamma(z)}$ represents the intrinsic opening angle, which decreases as $\Gamma(z)$ increases, leading to a transition from a parabolic to a conical jet shape. Eq.~\ref{Eq:R_z} relies on the small‐angle approximation $\tan(x)\approx x$\footnote{The intrinsic and apparent half-opening angles are related as $\tan(\eta_{\rm int}/2)=\tan(\eta_0/2)/ \sin \theta_{\rm obs} \rightarrow \eta_{\rm int} \approx \eta_0 /\Gamma $, assuming small angles and $\theta_{\rm obs} \sim 1/\Gamma$.}, which introduces $<5\%$ error for $x<0.4$ radians. We adopt $\eta_0=0.36$, consistent with the median apparent opening angle derived from a sample of 186 Fermi-LAT detected sources in the MOJAVE 1.5 Jy sample \citep{2017MNRAS.468.4992P} and the 11-month Fermi-LAT catalog \citep{2013AJ....146..120L}.

We assume that the jet plasma is composed of pairs and protons with number densities $n'_{\rm e}(z)$ and $n'_{\rm p}(z)$, respectively, as measured in the jet rest frame. The total number density can be expressed as $n'(z)=n'_{\rm p}(z)(1+k_{\rm p})$, where $k_{\rm p}$ is the pair multiplicity of the plasma without accounting for secondary pairs.

Assuming particle conservation along the jet we may write \citep{1979ApJ...232...34B}:
\begin{equation}
n'(z)= n'_0\left(\frac{\Gamma(z)\beta(z)}{\Gamma_{0}\beta_0}\right)^{-1}\left(\frac{R(z)}{R_0}\right)^{-2}.
\label{Eq:n_z}
\end{equation}
where $n'_0$ is the total number density of particles at the base of the jet and $\beta(z)$ is the bulk jet velocity at distance $z$ in units of $c$.

For a cold outflow, where the pressure and internal energy are
negligible compared to the rest mass energy of the plasma, the jet magnetization is defined as\footnote{The magnetization $\sigma$ refers to the unreconnected plasma at the jet location where dissipation takes place. However, magnetization of the reconnected plasma is close to unity~\citep[e.g.][]{2015MNRAS.450..183S, 2021ApJ...912...48H}.},

\begin{equation}
\sigma(z) = \frac{B'^2(z)}{4\pi n'_{\rm p}(z)m_{\rm p}c^2\left( 1 + \frac{m_{\rm e}}{m_{\rm p}}k_{\rm p} \right)} = 2\frac{U'_{\rm B}(z)}{U'_{\rm e}(z)+ U'_{\rm p}(z)}, 
\label{Eq:magnetization}
\end{equation}
where we introduced the magnetic field energy density $U'_{\rm B}(z)$, the pair energy density $U'_{\rm e}(z)$, and the proton energy density $U'_{\rm p}(z)$; these are equal to the rest-mass energies of each species in the cold plasma limit (low-$\beta$ plasma). Using Bernoulli's equation along a streamline of the flow, the total energy flux per unit rest-mass energy flux, $\mu$, is written as, 
\begin{equation}
\mu \equiv 
\Gamma_{0} \left(1 + \sigma_0 \right) = \Gamma_{\max} \left(1 + \sigma_{\rm acc} \right)
\label{Eq:Bernoulli}
\end{equation}
where $\sigma_0$ is the initial magnetization at the base of the jet and $\sigma_{\rm acc}$ is the magnetization at $z_{\rm acc}$. If $\sigma_{\rm acc} > 1$, then the conversion of the jet Poynting power to bulk kinetic power is not fully achieved by the distance $z_{\rm acc}$. By requiring the magnetization at $z_{\rm acc}$ to be positive, we determine the minimum initial magnetization $\sigma_0$ necessary to accelerate the jet to $\Gamma_{\max}$, 
\begin{equation}
\sigma_0> \frac{\Gamma_{\max}}{\Gamma_{0}}-1.
\label{Eq:min_sigma}
\end{equation}

The jet power can be written as \citep[e.g.][]{D02, Giannios2019}
\begin{equation}
L_{\rm jet} = L_{\rm k, jet} + L_{\rm P, jet} = \frac{1+\sigma}{\sigma} L_{\rm P, jet} = (1+\sigma) L_{\rm k, jet},
\label{eq:Ljet}
\end{equation}
where $L_{\rm P, jet}$ is the Poynting power of the jet (defined in Appendix~\ref{app_A}) and $L_{\rm k, jet} = \Gamma \dot{M}_{\rm out} c^2$ is the kinetic power of the jet. Here, $\dot{M}_{\rm out}$ is the mass outflow rate through the jet, which can be related to the mass accretion rate $\dot{M}_{\rm acc}$ as $\dot{M}_{\rm out}c^2 = \eta_{\rm j} \dot{M}_{\rm acc}c^2 = \eta_{\rm j} \dot{m} L_{\rm Edd}/\eta_{\rm c}$. The parameter $\dot{m}$ represents the dimensionless mass accretion rate, which is the ratio of the actual mass accretion rate to the Eddington accretion rate. $L_{\rm Edd}$ denotes the Eddington luminosity, expressed as $L_{\rm Edd} = 4 \pi G M c m_{\rm p}/\sigma_{\rm T} \simeq 1.3 \times 10^{47} \, M_9$~erg s$^{-1}$ and $M_9 \equiv M/10^9 M_{\odot}$. The quantity $\eta_{\rm c} \sim 0.1$ refers to the radiative efficiency of the accretion process. In terms of $\dot{m}, M$ and $\mu$ (defined in Eq.~\ref{Eq:Bernoulli}) the jet power can be expressed as
\begin{equation}
    L_{\rm jet} = \mu \frac{\eta_{\rm j}}{\eta_{\rm c}} \dot{m} L_{\rm Edd} \simeq 1.3 \times 10^{48}~{\rm erg \, s^{-1}} \, 
    \mu_{2} \dot{m}_{-1} M_{9} \eta_{\rm j, -1}, \eta_{\rm c,-1}^{-1},
\end{equation}
where we introduced the notation $q_x = q/10^x$.

We now proceed to compute the initial particle number density in the jet. We relate the total energy density and the pressure\footnote{We assume that the plasma is cold and that the magnetic pressure is $P'_B = U'_B$.} at the base of the jet to the jet power (Eq.~\ref{eq:Ljet}):

\begin{equation}
U'_{\rm e}(z_0) + U'_{\rm p}(z_0) +  2 U'_{\rm B}(z_0) = \frac{L_{\rm jet}}{\pi R_0^2 \Gamma_{0}^2 \beta_0 c}.
\label{Eq:Init_en_dens}
\end{equation}

By using Eqs.~\ref{Eq:magnetization} and \ref{Eq:Init_en_dens} we can compute the total particle number density at the base of the jet,
\begin{eqnarray}
n'_0 & =&\frac{L_{\rm jet}}{\pi R_0^2 \Gamma_{0}^2 \left(1+\sigma_0\right) \beta_0 c }\frac{1+k_{\rm p}}{m_{\rm p} c^2\left(1+k_{\rm p}\frac{m_{\rm e}}{m_{\rm p}} \right)}  \\
&\simeq & 10^{8.3}~{\rm cm^{-3}} \,  \frac{L_{\rm jet, 47}}{\sigma_{0.1}}  \frac{1+k_{\rm p}}{1+k_{\rm p}\frac{m_{\rm e}}{m_{\rm p}}},
\label{Eq:Init_num_density}
\end{eqnarray}
where the numerical value was obtained for $R_0 = 3 R_{\rm s}$ with $R_{\rm s}$ being the Schwarchild radius and $\Gamma_{0} = 1.1$. 

Similarly, we may estimate the magnetic field strength at the jet base\footnote{The same equation also yields the strength of the comoving magnetic field at distance $z$ if all quantities with subscript '0' are replaced by their values at $z$.},
\begin{equation}
B'_0 =\left( \frac{4 \sigma_0 L_{\rm jet}}{(1+\sigma_0) R_0^2\Gamma_0^2 \beta_0 c }\right)^{1/2} \simeq 6 ~{\rm kG} \,  L_{\rm jet, 47}^{1/2},
\label{Eq:B0}
\end{equation}
where the numerical value was obtained for $R_0 = 3 R_{\rm s}$, $\Gamma_{0} = 1.1$,  and assuming that $k_{\rm p}\ll m_{\rm p}/m_{\rm e}$. Notice that the magnetic field strength is independent of the jet magnetization, if $\sigma_0 \gg 1$. 

\subsection{Energy dissipation and particle acceleration}\label{Sec:Particle_distribution}
Charged particles accelerated by reconnection obtain power-law distributions with index $p\equiv -{\rm d}\log N'/{\rm d}\log \gamma '$ depending on magnetization, as demonstrated by particle-in-cell (PIC) simulations of reconnection in electron-positron \citep{2014ApJ...783L..21S,2014PhRvL.113o5005G} and ion-electron plasmas \citep{2016ApJ...818L...9G}.
For simplicity, we assume that the slope of the protons is the same as the slope of pairs $p=p_{\rm e}=p_{\rm p}$. We used the results of \cite{Ball2018} (proton and electron plasma) for $\sigma<1$ and \cite{Werner2016} (pair plasma) for $\sigma>1$ to estimate the power-law index of the injected particles. These results were obtained for low plasma parameters $\beta$ (ratio between proton thermal pressure and magnetic pressure), which is consistent with the plasma being cold. 
\begin{table}
\centering
\begin{tabular}{cc}
\hline
Coefficient & Value \\
\hline
$a_6$ & 0.059452 \\
$a_5$ & -0.434136 \\
$a_4$ & 0.962786 \\
$a_3$ & -0.315536 \\
$a_2$ & -0.833513 \\
$a_1$ & -0.388311 \\
$a_0$ & 2.584913 \\
\hline
\end{tabular}
\caption{Coefficients of the 6th-order polynomial fit to the combined $\log_{10}(\sigma)$ vs power-law index data. The fit is of the form $f(x) = \sum_{i=0}^{6} a_i x^i$, where $x = \log_{10}(\sigma)$.}
\label{tab:polyfit_coefficients}
\end{table}

\begin{figure}
\centering
\includegraphics[width=0.4\textwidth]{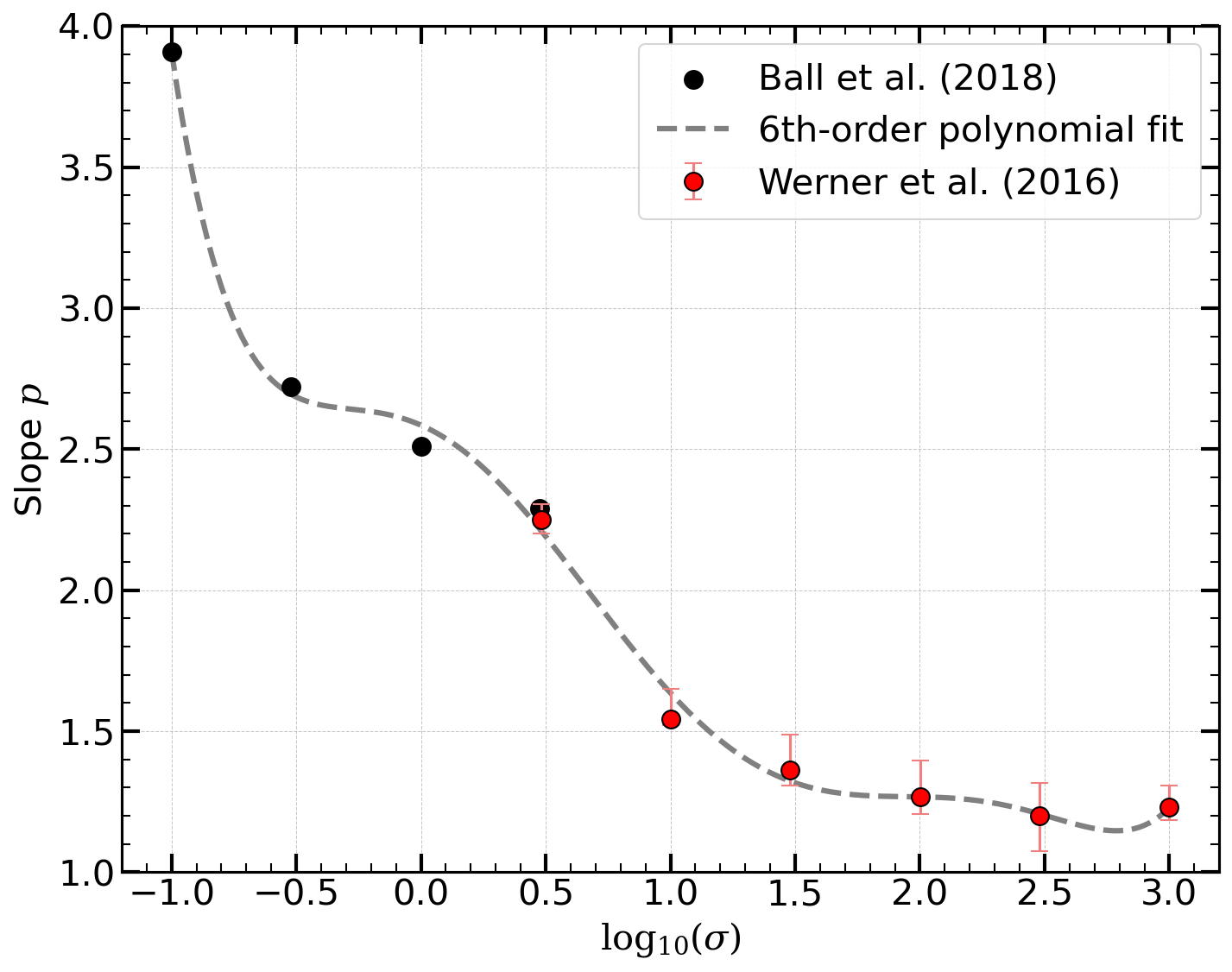} 
\caption{Power-law index $p$ of the injected particles distribution
as a function of the magnetization $\sigma$.}
\label{fig:sigma_vs_p}
\end{figure}

\begin{figure}
\centering
\includegraphics[width=0.4\textwidth]{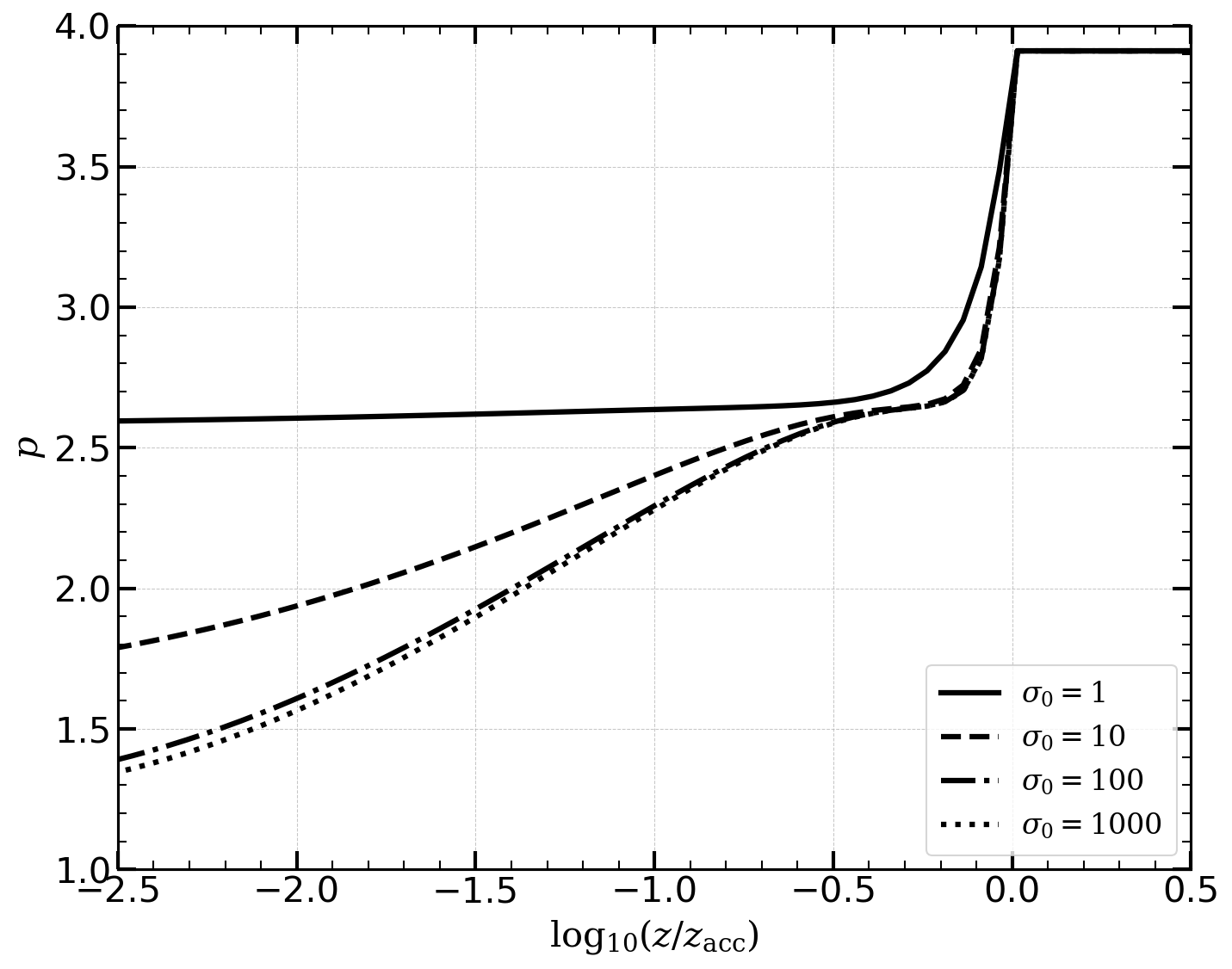} 
\caption{Power-law index $p$ of the injected particle distribution in the jet as a function of distance $z$ from the SMBH. Coloured curves correspond to different values of the initial jet plasma magnetization $\sigma_0$. Other parameters used for this plot are $ \Gamma_{0}=1.1, \Gamma_{\rm max}=\Gamma_{0}(1+\sigma_0)/\Gamma_0, M = 10^9M_{\odot}, z_{\rm acc} = 10^3~R_{\rm s}$.}
\label{fig:z_vs_p}
\end{figure}

In Fig.~\ref{fig:sigma_vs_p}, we present the power-law index as a function of $\sigma$. We observe a continuous change at $\sigma \sim 3$ where the two datasets overlap. To ensure a smooth transition, we fit these data points with a sixth-order polynomial: $f(\sigma)=a_0+a_1\sigma+a_2\sigma+...+a_6\sigma$. The best-fit coefficients are listed in Table~\ref{tab:polyfit_coefficients}. For $\sigma > 10^3$, we set $p=1.2$. In Fig.~\ref{fig:z_vs_p} we present the power-law index as a function of distance from the SMBH for several values of initial magnetization $\sigma_0$. We make the choice to set the asymptotic value of the bulk Lorentz factor as $\Gamma_{\rm max} = \Gamma_{0}(1+\sigma_0)/1.1$ which leads to $\sigma_{\rm acc} = 0.1$ since this is the lower value of magnetization so we can estimate the power law index. Since the bulk acceleration of the jet is driven by magnetic fields, larger values of $\sigma_0$ (indicating a higher ratio of magnetic energy density to particle energy density) result in higher magnetizations at the location of the dissipation, thus leading to generally harder energy distributions of the accelerated particles. For small values of $\sigma_0$, the acceleration is insufficient to achieve the high terminal Lorentz factors found for larger $\sigma_0$ (see Eq.~\ref{Eq:min_sigma}).

We assume that the energy injection rate in relativistic particles is isotropic in the jet frame, and that the accelerated particle distributions receive a fraction $f_{\rm rec}$ of the jet Poynting luminosity $L_{\rm P,jet}$ (see Eq.~\ref{Eq:Inj_lum}). The volumetric injection rate of relativistic particles by reconnection is given by,
\begin{equation}
    \frac{{\rm d}N'_{\rm i}}{{\rm d}V' {\rm d}\gamma' {\rm d}t'}=Q^{\rm inj}_{\rm i}(\gamma) = Q_{\rm 0,i}\gamma'^{-p(\sigma)},\ \ \  \gamma'_{\rm i,min}<\gamma'<\gamma'_{\rm i,max}
\label{Eq:Inj_rate}
\end{equation}
where $\gamma'_{\rm i,min}, \gamma'_{\rm i,max}$ are the minimum and the maximum particle Lorentz factors, and $Q_{\rm 0,i}$ is a normalization factor, which can be derived by equating $m_{\rm i} c^2 V' \int {\rm d}\gamma \, \gamma Q_{\rm i}^{\rm inj}(\gamma)$ with Eq.~\ref{Eq:Inj_lum}. Here, $V' = 4 \pi R^3/3$ is the volume of the emitting region, which is assumed to be spherical in the jet-comoving frame. 

The minimum Lorentz factor is set to unity for $p<2$ (or $\sigma>5$), as it is a mundane parameter. However, for $p>2$ (or $\sigma < 5$) we follow \cite{Petro2023}
and define it as, 
\begin{equation}
   \gamma'_{\rm i,min}=\frac{2-p}{1-p}f_{\rm rec}\sigma \frac{m_{\rm p}}{m_{\rm i}}.
\label{Eq:gamma_min}
\end{equation}
The maximum available energy that can be achieved by a particle with charge $q$ in the reconnection region is given from,
\begin{equation}
    E_{\rm i,max}(z) = qE'_{\rm rec}(z)R(z)
\label{Eq:HC}
\end{equation}
where $E'_{\rm rec}$ is the ideal electric field in the reconnection region, $E'_{\rm rec}(z) \sim \beta_{\rm rec} B'(z)$, and $\beta_{\rm rec}\sim 0.06$ is the reconnection rate. Charged particles can reach such energies only if the acceleration process is faster than the advection of the particles from the system, and the energy losses are not relevant. Assuming that the acceleration is driven by the non ideal electric field in the reconnection region, we can write the acceleration rate as follows,

\begin{equation}
    t_{\rm acc} \equiv \frac{\gamma'}{\dot{\gamma}'_{\rm acc}}= \frac{\gamma' m_{\rm i}c}{\beta_{\rm rec} q B'(z)} = \eta_{\rm acc} \frac{r_{\rm g}}{c}.
\label{Eq:t_acc}
\end{equation}
where $r_{\rm g}$ is the gyroradius of a relativistic particle and $\eta_{\rm acc} = 10 \, \beta^{-1}_{\rm rec, -1}$ can be thought of as an acceleration efficiency. Indeed, recent three-dimensional PIC simulations of reconnection have found $\eta_{\rm acc}\sim 10$ \citep{Zhang_2021, Zhang_2023}. To account for slower acceleration, we set $\eta_{\rm acc}= 10^4$  (see also \cite{Petro2023}), and demonstrate the impact of our choice on the jet emission in Sec.~\ref{sec:eta_acc}.

\begin{figure}
\centering
\includegraphics[width=0.45\textwidth]{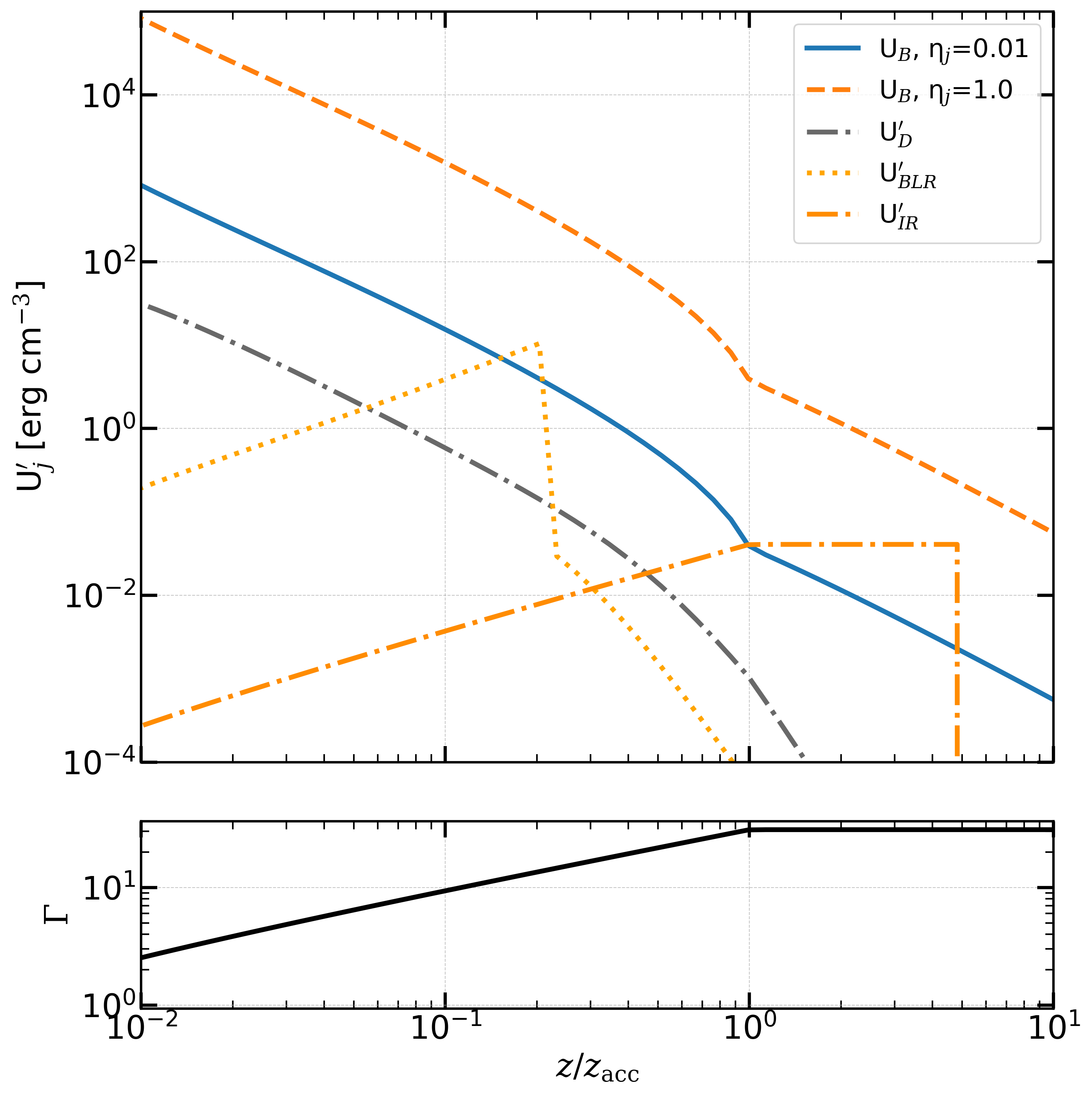} 
\caption{Evolution of energy densities in the comoving frame (top panel) and Lorentz factor (bottom panel) as functions of normalized distance $z/z_{\rm acc}$ for two values of $\eta_{\rm j}$, for $\dot{m}=0.01$ and for $\sigma_0=30$. The parameters used for this plot are $ \Gamma_{0}=1.1, \Gamma_{\rm max}=31, M = 10^9\, M_{\odot}, z_{\rm acc} = 10^3 R_{\rm s}$, and $k_{\rm p}=10$.}
\label{fig:Energy_dens_vs_z}
\end{figure}

\begin{figure}
\centering
\includegraphics[width=0.45\textwidth]{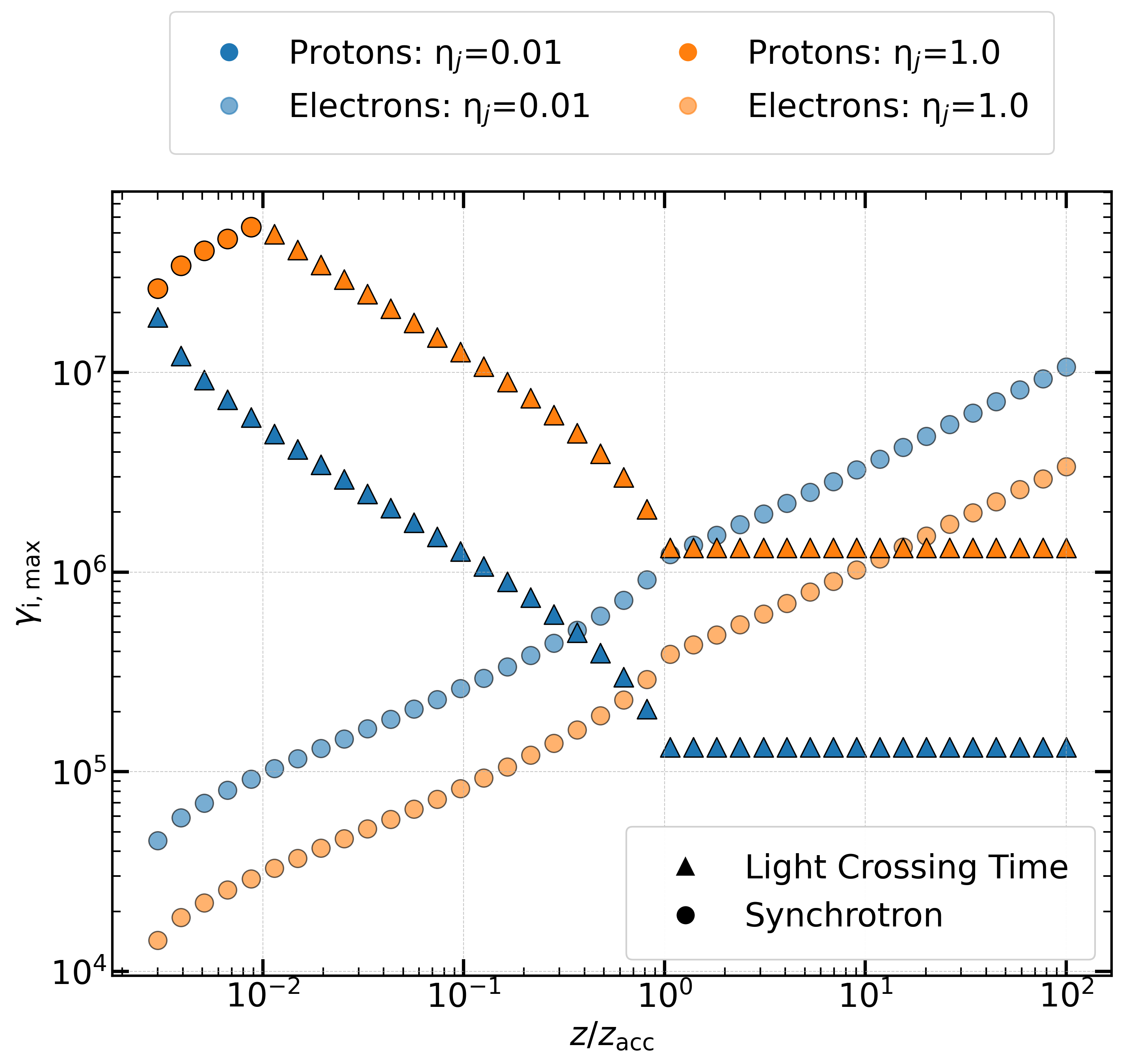} 
\caption{Maximum Lorentz factor of pairs and protons as functions of normalized distance $z/z_{\rm acc}$ from the SMBH for magnetization parameters $\sigma_0=30$, for $\dot{m}=0.01$ and two values of $\eta_{\rm j}$. Markers indicate the dominant processes constraining acceleration.  The parameters used for this plot are $ \Gamma_{0}=1.1, \Gamma_{\rm max}=31, M = 10^9M_{\odot}, z_{\rm acc} = 10^3 R_{\rm s}$, and $k_{\rm p}=10$.}
\label{fig:min_gamma_vs_z}
\end{figure}

\subsection{Radiative energy losses}\label{Subsec:Radiative_losses}
Pairs primarily lose energy through synchrotron radiation and IC scattering of low-energy photons. Protons, in addition to the two previous nonthermal processes, can lose energy via Bethe-Heitler pair production and photopion production when interacting with radiation. The maximum Lorentz factor $\gamma'_{\rm i,max}$ attained by pairs and protons is determined by the minimum of Eq.~\ref{Eq:HC} and the radiation-limited Lorentz factor $\gamma'_{\rm rad}$  (where the acceleration rate equals the energy loss rate), 
\begin{equation}
    \gamma'_{\rm i, max} = \min \left(\frac{qE'_{\rm rec} R}{m_{\rm i}c^2}, \gamma'_{\rm rad} \right)
    \label{Eq:gamma_max}
\end{equation}

The photon field in the dissipation region of the jet consists of two components: external thermal radiation originating from the accretion disc, the BLR, and the IR dusty torus (see Appendix~\ref{app:external}), and non-thermal radiation produced by relativistic particles. The latter requires modelling of the emitting region. For the magnetic field therein, we adopt the magnetic field of the unperturbed jet at the same distance, as provided by Eq.~\ref{Eq:magnetization}. In Fig.~\ref{fig:Energy_dens_vs_z} we present the comoving energy densities of the magnetic field and the external radiation fields (top panel) and the bulk Lorentz factor (bottom panel) as a function of the normalized distance $z/z_{\rm acc}$ for two cases of $\eta_{\rm j}$, $\dot{m}=0.01$ and $\sigma_0=30$. For this example, we choose $z_{\rm acc}=10^3\, R_{\rm s}$ and $\Gamma_{\rm max}=31$. For these choices of $\dot{m}$ and $\sigma_0$ we find that the energy density of the magnetic field, which scales as $B'^2\propto z^{-2}$, is dominant in the system. Because the energy density of the magnetic field is $U'_{\rm B}\propto \eta_{\rm j}\dot{m}$ we find that for $\eta_{\rm j}=0.01$ the comoving BLR energy density dominates the energy densities for distances close to $R_{\rm BLR}$, where the bulk Lorentz factor reaches its asymptotic limit.  In Fig.~\ref{fig:min_gamma_vs_z} we present the maximum attainable Lorentz factor of protons and pairs for two values of $\eta_{\rm j}$. We find that the acceleration of pairs is dictated by synchrotron losses, while the maximum Lorentz factor that they attain is an increasing function of distance because of the drop in the magnetic field across the distance of the jet. The maximum proton Lorentz factor, besides the case of high $\eta_{\rm j}$, which translates into higher magnetic fields across the jet, is driven by $t_{\rm acc}=t_{\rm cr}$ where $t_{\rm cr}$ is the light crossing time of the emitting region. 

We can estimate the distance at which the BLR energy density exceeds that of the accretion disc in the jet comoving frame by using the analytical approximation from \cite{dermer2009high} (see Eq.~6.153)\footnote{This expression holds as long as the distance of the emitting region is smaller than the BLR radius and the outer disc radius.}
\begin{equation}
r^{*}\simeq 10^{17}\left(\dot{m}_{-1}\,\ M_9^2\right)^{1/3}~{\rm cm}, 
\label{Eq:z_BLR_AD}
\end{equation}
which in terms of $z_{\rm acc}=10^3 R_{\rm s}$ reads $r^*/z_{\rm acc} \approx (1/3) \dot{m}_{-1}^{1/3} M_9^{-1/3}$ -- see also Fig.~\ref{fig:Energy_dens_vs_z}. 
 
We close this section by summarizing the key model parameters (for a description of all parameters, see Table~\ref{tab:param}). These are the dimensionless mass accretion rate $\dot{m}$, which controls the luminosity of the accretion disc, the size of the BLR and, the jet efficiency factor $\eta_{\rm j}$, which, together with $\dot{m}$, sets the energy budget available for jet formation and acceleration, the initial magnetization $\sigma_0$, which controls the energy partition between non-thermal particles and magnetic fields and governs the jet terminal Lorentz factor, and the particle acceleration efficiency $\eta_{\rm acc}$ that controls the maximum energy achieved by the non-thermal particles. 

\renewcommand{\arraystretch}{1.1}

\begin{table*}
    \caption{Model parameters with their description and values.}
    \centering
    \begin{threeparttable}
    \begin{tabular}{lcccc}
     \hline 
     Parameter     &  Symbol & Value(s) \\
     \hline
     Input (varying)\tnote{*} \\
     \hline
     Initial magnetization   &  $\sigma_0$  & 30 (10, 60) \\
     Mass accretion rate (in units of Eddington accretion rate) & $\dot{m}$ & 0.01 (0.001, 0.1) \\
     Ratio of jet power to accretion power & $\eta_{\rm j}$ & 0.1 (0.01, 1) \\
     Particle acceleration efficiency & $\eta_{\rm acc}$ & $10^4$ ($10^3$, $3\cdot 10^4$, $10^6$) \\ 
     \hline
     Input fixed \\
     \hline  
     Black hole mass & $M_{\rm BH}$ & $10^9 M_{\odot}$ \\
     Initial jet Lorentz factor & $\Gamma_0$ & 1.1 \\
     Dissipated energy fraction transferred to relativistic particles & $f_{\rm rec}$ & 0.25 \\    
     Minimum electron Lorentz factor & $\gamma_{\rm e,\min}$ & $10^{0.1}$ (for $p<2$) \\
     Minimum proton Lorentz factor & $\gamma_{\rm p,\min}$ & $10^{0.1}$ (for $p<2$) \\ 
     Pair-to-proton number ratio & $\kappa_{\rm p}$ & 10 \\
     Disc radiative efficiency & $\eta_{\rm c}$ & 0.1 \\
     BLR reprocessing fraction & $f_{\rm BLR}$ & 0.1 \\
     DT reprocessing fraction & $f_{\rm DT}$ & 0.1 \\
     Distance of jet bulk acceleration & $z_{\rm acc}$ & $10^3 R_{\rm s}$ \\
     Observation angle between the jet axis and the observer’s line of sight & $\theta_{\rm obs}$ & $2~\rm deg$\\
     \hline 
     Derived \\
     \hline 
     Terminal jet Lorentz factor & $\Gamma_{\max}$ & Eq.~\ref{Eq:Bernoulli} \\
     Doppler factor & $\delta$ &  Eq.~\ref{Eq:delta} \\
     Blob radius & $R'$~(cm) & Eq.~\ref{Eq:R_z} \\
     Magnetic field strength of unreconnected plasma\tnote{$\ddag$} & $B'$~(G) & Eqs.~\ref{Eq:n_z}-\ref{Eq:magnetization}\\
     Total jet power & $L_{\rm jet}$~(erg s$^{-1}$) & Eq.~\ref{eq:Ljet} \\
     Injection luminosity of particle species $i$ & $L'_{\rm i}$~(erg s$^{-1}$) & Eq.~\ref{Eq:Inj_lum} \\
     Power-law slope of particle distribution & $p$ & Polynomial fit (see Fig.~\ref{fig:sigma_vs_p}) \\
     Minimum Lorentz factor of particle species $i$ & $\gamma_{\rm i,\min}$ & Eq.~\ref{Eq:gamma_min} (for $p\ge 2$) \\
     Maximum Lorentz factor of particle species $i$ & $\gamma_{\rm i,\max}$ & Eq.~\ref{Eq:gamma_max} \\    
     \hline     
    \end{tabular}
    \begin{tablenotes}
\item[*] The values outside the parenthesis correspond to the baseline model. The values inside the parenthesis are those considered in the parametric study. 
\end{tablenotes}
\end{threeparttable}
    \label{tab:param}
\end{table*}

\section{Methods}\label{sec:methods} 

We assume that the non-thermal charged particles and photons occupy a spherical region (blob) whose radius $R'_{\rm b}$ coincides with the cross section of the jet at a given distance. The magnetic field in this blob is assumed to be uniform and is the same as the magnetic field of the unperturbed flow at the same distance $z$ (see Eq.~\ref{Eq:magnetization}). The evolution of the non-thermal particles involves the solution of a system of differential equations (kinetic equations). The kinetic equations for a homogeneous emitting region containing relativistic particles of species $i$ can be written as follows,

\begin{equation}
\begin{split}
    \frac{\partial N'_{\rm i}(\gamma'_i,t')}{\partial t'} 
    + \frac{N'_{\rm i}(\gamma'_i,t')}{ t'_{\rm esc,i}} 
    + \sum_{\rm j} \mathcal{L}_{\rm i}^{\rm j}(N'_{\rm i},N'_{\rm k},t')= \\
    = \sum_{\rm j} Q_{\rm i}^{\rm j}(N'_{\rm i},N'_{\rm k},t') 
    + Q^{\rm inj}_{\rm i}(\gamma'_{\rm i},t')
\end{split}
\label{Eq:kin_eq}
\end{equation}
where $N'_{\rm i}$ is the differential number of the non-thermal particles, $t'_{\rm esc,i}$ is the escape timescale of the non-thermal particles which we set equal to the light crossing time of the emitting region $R'_{\rm b}/c$. The energy loss term for non-thermal particles due to the $j$ process is denoted $\mathcal{L}_{\rm i}^{\rm j}$, while $Q_{\rm i}^{\rm j}$ is the operator for non-thermal injection due to the $j$ process and $Q^{\rm inj}_{\rm i}$ is the external injection of particles into the blob (primaries). While modeling the source we account for pair and proton synchrotron emission and absorption, IC scattering (for pairs only), $\gamma \gamma$ annihilation, photopair (Bethe–Heitler pair production), and photopion production processes. To solve the coupled kinetic equations of the non-thermal particles in the blob namely, electrons (and positrons), protons, photons, and neutrinos (and antineutrinos) we utilized {\tt LeHaMoC} \citep{2024A&A...683A.225S}, an open-source time-dependent leptohadronic modeling code. We compute the (comoving) photon and all-flavor neutrino energy spectra until an equilibrium is reached (steady state) since we are not interested in studying transient phenomena such as blazar flares. Finally, we perform the appropriate transformations to obtain the spectra in the observer’s frame using the Doppler factor.

\section{Results}\label{sec:results}
In this section we present numerical results for the jet model (Sec.~\ref{sec:jet-model}) and its non-thermal emission for leptonic scenarios (Sec.~\ref{sec:leptonic}) and leptohadronic scenarios (Sec.~\ref{sec:leptohadronic}).

\begin{figure*}
\centering
\includegraphics[width=0.95\textwidth]{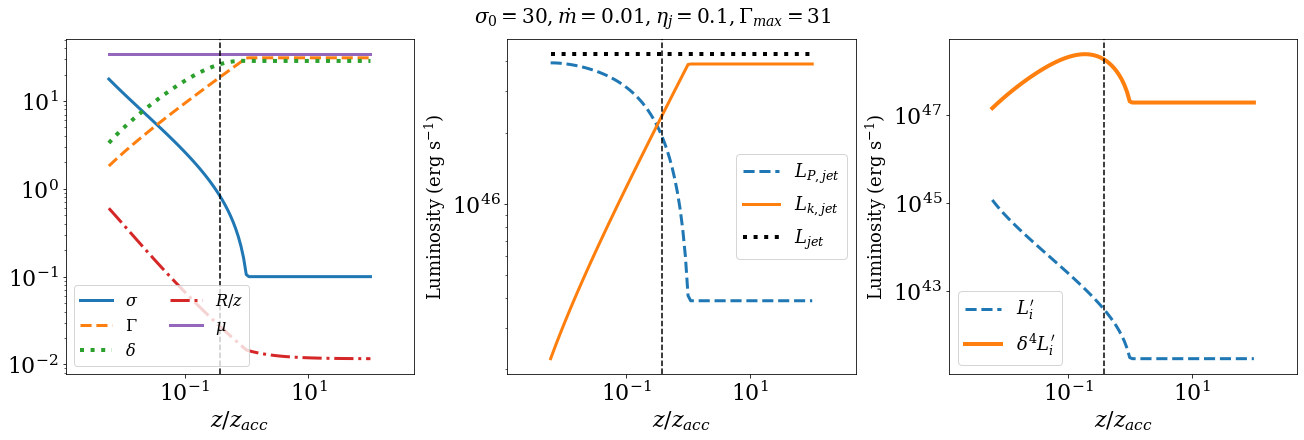} 
\caption{Evolution of jet properties and characteristic luminosities with distance from the central engine for $\sigma_0 = 30$, $\dot{m} = 0.01$, and $\Gamma_{\rm max}=31$. All other parameters are listed in Table~\ref{tab:param}. \textit{Left panel:} Evolution of the bulk Lorentz factor $\Gamma$, the magnetization $\sigma$, the Doppler factor $\delta$, and the emitting region size $R'$ normalized to the distance $z$ as functions of $z/z_{\rm acc}$. \textit{Middle panel:} Poynting jet power $L_{\rm P,jet}$ and kinetic jet power $L_{\rm k, jet}$ as a function of normalized distance $z/z_{\rm acc}$. The sum of the two components yiedls the total jet power $L_{\rm jet}$. \textit{Right panel:} Comoving particle injection luminosity $L'_{\rm i}$ and its respective value in the observer's frame $\delta^4 L'_{\rm i}$ as a function of the normalized distance $z/z_{\rm acc}$ from the SMBH. The vertical dashed line in all panels marks the distance of the BLR, which is close to $z_{\rm acc}$ for the baseline model parameters. }
\label{fig:Parameters_vs_z}
\end{figure*}

\subsection{Jet model}\label{sec:jet-model}
We present in Fig.~\ref{fig:Parameters_vs_z} the evolution of bulk jet properties, such as plasma magnetization, Lorentz factor, and various luminosities, with distance from the SMBH for the baseline model parameters: $\dot{m} = 0.01,\quad \sigma_0 = 30,\quad \eta_{\mathrm{acc}} = 10^4,\quad z_{\rm{acc}} = 10^3\,R_{\rm s},\quad \eta_{\rm j} = 0.1$ for other model parameters, see Table~\ref{tab:param}. In the left panel of Fig.~\ref{fig:Parameters_vs_z} we present the evolution of plasma magnetization $\sigma$ (solid blue line), bulk Lorentz factor $\Gamma$ (dashed orange line), Doppler factor $\delta$ (dotted green line), and the ratio $R/z$ (dot-dashed red line) as functions of $z/z_{\rm acc}$. The total energy per baryon $\mu$ (solid purple line) remains constant throughout the jet evolution. The vertical dashed line marks in all panels correspond to the location of the BLR, which is close to $z_{\rm acc}$. The jet undergoes gradual acceleration, with $\Gamma$ increasing and $\sigma$ decreasing as the energy is converted from magnetic to kinetic form. However, we choose to set $\sigma_{\rm acc}=0.1$ since this is the lower value of magnetization so we can have an estimation for the power law index of the nonthermal particles (see also Fig.~\ref{fig:sigma_vs_p}). In the middle panel of Fig.~\ref{fig:Parameters_vs_z} we present the evolution of the Poynting flux luminosity $L_{\rm P,jet}$ (dashed blue line), kinetic luminosity $L_{\rm k,jet}$ (solid orange line), and total jet power $L_{\rm jet}$ (dotted black line) as functions of $z/z_{\rm acc}$. At small distances, $L_{\rm P,jet}$ dominates over $L_{\rm k,jet}$, but as the jet accelerates, the energy is gradually transformed into bulk energy. In the right panel of Fig.~\ref{fig:Parameters_vs_z} we present the comoving injection luminosity $(L'_{\rm i}$ to the non-thermal particles, and the observed injected luminosity $\delta^4 L'_{\rm i}$ as functions of the normalized distance $z/z_{\rm acc}$ from the SMBH. For distances $z > z_{\rm acc}$, the comoving injection luminosity $L'_{\rm i}$ remains constant because $L_{\rm P,jet}$ is constant and the bulk Lorentz factor saturates at its asymptotic value. The observed injected luminosity $\delta^4 L'_{\rm i}$ attains its maximum at $z/z_{\rm acc} \sim 10^{-1}$, corresponding to the distance where $\Gamma \sim 1/\theta_{\rm obs}$, maximizing the Doppler factor $\delta$. 

\subsection{Leptonic model} \label{sec:leptonic}

\begin{figure}
\centering
\includegraphics[width=0.45\textwidth]{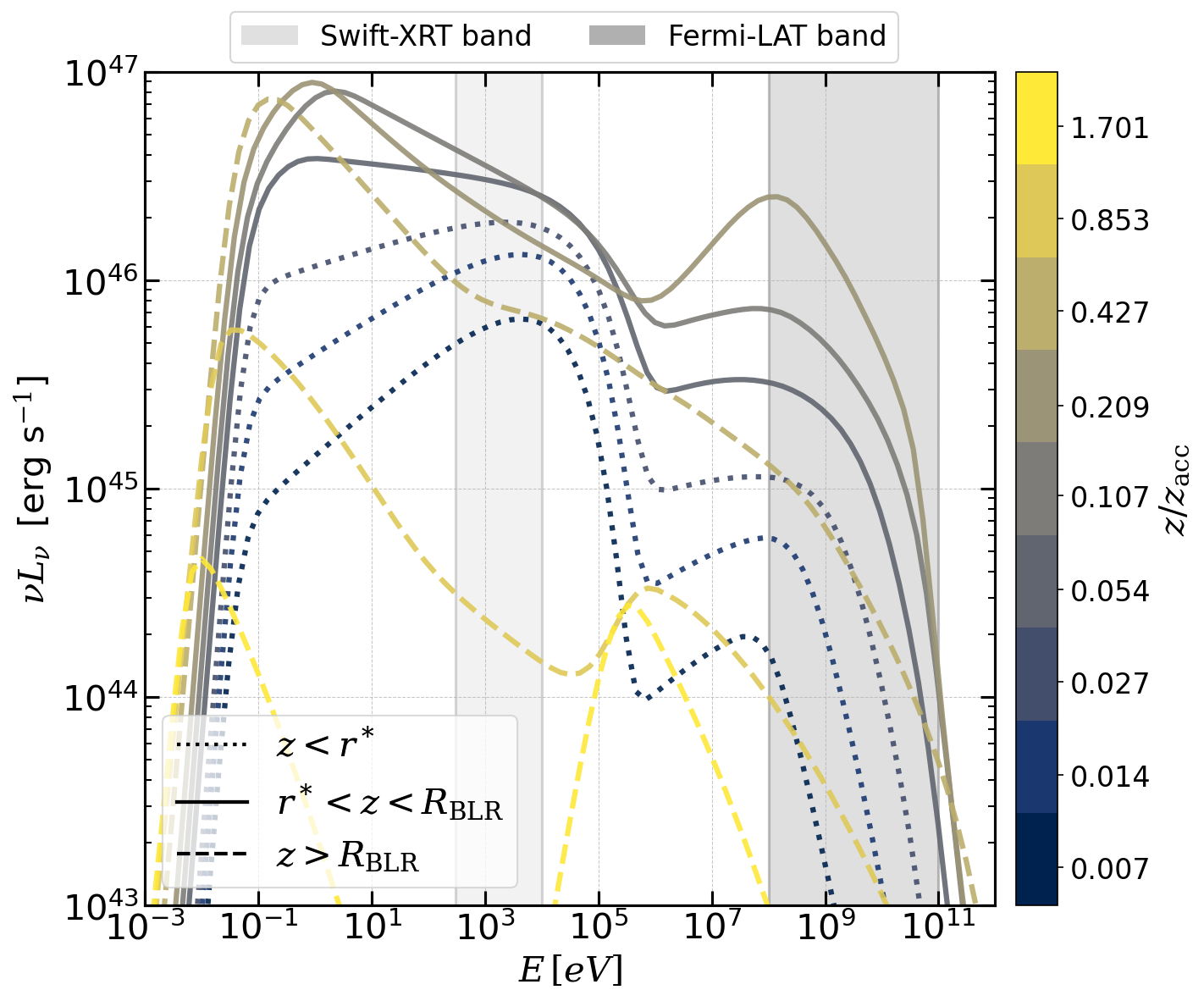} 
\caption{SEDs for the baseline model as a function of distance from the SMBH. Solid lines represent SEDs produced by emitting regions at $z < R_{\mathrm{BLR}}$ while dashed lines represent SEDs with emitting regions located beyond the BLR. The shaded light gray and darker gray bands highlight the 0.3–10~keV and 100 MeV–100 GeV energy ranges, respectively.}
\label{fig:baseline_vs_z}
\end{figure}

\subsubsection{Baseline model}
In Fig.~\ref{fig:baseline_vs_z} we present the photon SEDs calculated for the baseline model assuming different locations of the emitting region. The emission is boosted to the observer's frame according to the Lorentz factor of the location of the unperturbed flow (see Eq. \ref{Eq:G_z}). Different colours indicate the normalized jet distance $z / R_{\rm BLR}$, following the colour scale on the right of the plot. Different line styles are used to mark different locations of the emitting region along the jet. The shaded areas indicate observational energy bands of interest\footnote{The selected energy ranges correspond to Swift-XRT and Fermi-LAT that play a key role in blazar observations.}, namely 0.3-10~keV (light gray) and 100 MeV -- 100 GeV (darker gray). In the SEDs presented, we do not account for absorption due to extragalactic background light (EBL). In this plot, we can identify different families of SEDs depending on the distance from the SMBH. These are determined by the dominant radiative process at a given distance from the SMBH and thus are related to the energy densities of the magnetic field and the external fields of the AD, the BLR, and the IR DT. For $z<r^{*}$ (see Eq.~\ref{Eq:z_BLR_AD}) in which the dominant external photon field is the AD, we find that most luminosity is emitted by pairs through synchrotron radiation, since $U'_{\rm B} > U'_{\rm D}$ as shown in Fig.~\ref{fig:Energy_dens_vs_z}. The high-energy component is attributed to the synchrotron self-Compton process. For $r^{*} < z < R_{\rm BLR}$, the dominant source of external photons is the BLR. Its energy density is boosted in the rest frame of the emitting region, enhancing the rate of IC scattering and resulting in a more prominent high-energy component. Beyond the BLR, the photon density of the BLR is deboosted, and the dominant source of external photons is the photons of the IR DT. The high-energy emission in the $\gamma$-rays drops because of the decreasing magnetic field (SSC emission), change in particle distribution shape, and injected luminosity $L'_{\rm e}$ (see right panel of Fig~\ref{fig:Parameters_vs_z}). The complete SED is shaped by synchrotron and EC processes, with the emission of the latter being at the $\sim$MeV energy band. The three families of SEDs are also characterised by different bolometric luminosities, with those produced in the narrow range between $r^*$ and $R_{\rm BLR}$ being the most luminous. This trend is the combined result of how the radiative efficiency and the injected power into particles depend on the distance from the SMBH.

\begin{figure*}
\centering
\includegraphics[width=0.95\textwidth]{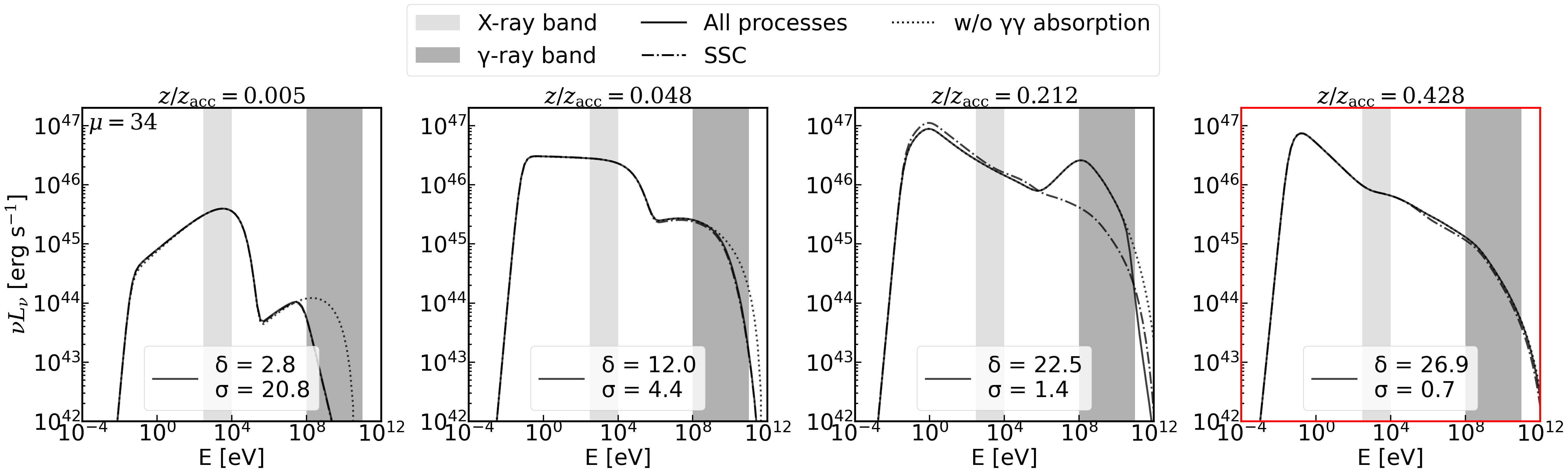} 
\includegraphics[width=0.95\textwidth]{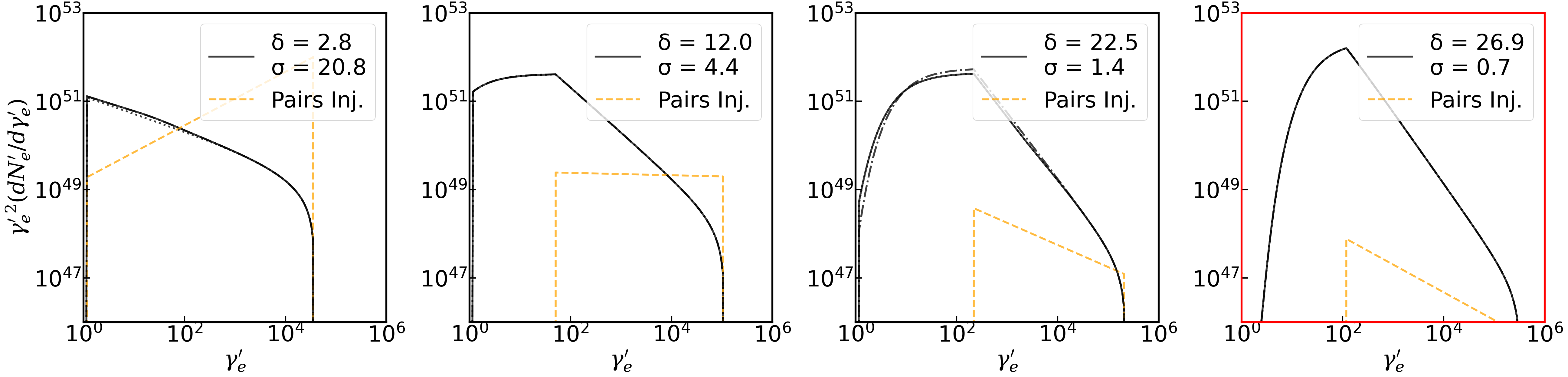} 
\caption{\textit{Top row:} SEDs at different distances $z$ from the SMBH for our baseline model. The dash-dotted lines represent emission from primary pairs without external photon fields (SSC-only case), while the dotted lines show spectra without $\gamma\gamma$ absorption.  All other parameters are fixed to the values provided in Table~\ref{tab:param}. Black solid borders indicate an emission region located inside the BLR, while the red solid borders indicate an emission region located beyond the BLR. The shaded gray regions correspond to the Swift-XRT (light gray) and Fermi-LAT (darker gray) energy bands. \textit{Bottom row:} Steady-state pair distributions at different distances $z$ from the SMBH for our baseline model. The orange dashed lines show the differential pair distribution at injection.}
\label{fig:SEDs_decomp}
\end{figure*}

To better understand the influence of external photon fields and the resulting EC emission on the high-energy component of the SED, we present Fig.~\ref{fig:SEDs_decomp}. The top row shows the SEDs, while the bottom row presents the corresponding steady-state comoving electron–positron distributions for the baseline model. Each column corresponds to a different distance $z$ from the SMBH, where a different external photon field as seen in the emitting region is dominant. The bottom row clearly illustrates how the particle distribution depends on magnetization $\sigma$. The injected electron spectra (see orange dashed lines in Fig.\ref{fig:SEDs_decomp}) become softer as magnetization decreases with distance. Specifically, if the emitting region is formed at a high magnetization ($\sigma > 5$) region of the jet, the accelerated particles attain harder power laws. Notably, at $z \approx \,R_{\rm BLR}$ (third column), EC losses slightly reshape the electron distribution (solid lines) compared to the SSC solution (dash-dotted lines) due to enhanced cooling from the BLR photon field. The additional source of cooling extends the pair distribution toward lower energies relative to the SSC-only scenario. 

In the top row, the dash-dotted lines represent the emission produced by the primary pairs without external photon fields, while the dotted lines represent the spectrum without $\gamma \gamma$ absorption. For regions where $\sigma > 5$, the injected non‐thermal particles follow a hard power-law ($p < 1.5$), leading to a synchrotron peak around $\sim 10\,\mathrm{keV}$. Most of the radiative output in these cases is produced by high-energy pairs near $\gamma_{\rm e} \sim \gamma_{\rm e, max}$. For distances at which $ \sigma < 5$, the injected particle spectrum and SEDs are softer.  

Overall, synchrotron emission dominates at nearly all distances, both with and without external fields. However, at $z \approx \,R_{\rm BLR}$ (third column), the boosted BLR photons enhance the EC emission by up to a factor of $\sim 5$ over the SSC component (compare the solid and the dash‐dotted line in the third column of Fig.~\ref{fig:SEDs_decomp}), thus producing prominent high-energy peaks in the $\gamma$-ray range with a power-law spectrum in the Fermi-LAT band (100 MeV -- 100 GeV). We also observe that $\gamma$-ray absorption is only significant at $z < R_{\rm{BLR}}$, where the density of external photons is relevant. The cascade emission at lower energies is negligible in our baseline model (compare the dotted and solid lines in Fig.~\ref{fig:SEDs_decomp}).  

Beyond the BLR ($z > R_{\rm BLR}$), indicated by the plot with the red highlighted borders in Fig.~\ref{fig:SEDs_decomp}, the external field is deboosted and thus can be neglected, leaving synchrotron radiation as the dominant emission process. At these distances, the contribution of the IR DT photon field is negligible compared to the magnetic field energy density (see Fig.~\ref{fig:Energy_dens_vs_z}). Finally, the radiated luminosity follows the luminosity injected into the particles (dashed line in the right panel of Fig.~\ref{fig:Parameters_vs_z}) and peaks at $z / z_{\mathrm{acc}} \sim 0.1$.

In the following sections, we systematically explore how changes in the mass accretion rate $\dot{m}$, initial magnetization $\sigma_0$, particle acceleration efficiency $\eta_{\rm acc}$, and $z_{\rm acc}$  modify the resulting SEDs at various distances in the jet for two values of $\eta_{\rm j}$.

\begin{figure*}
\centering
\includegraphics[width=0.95\textwidth]{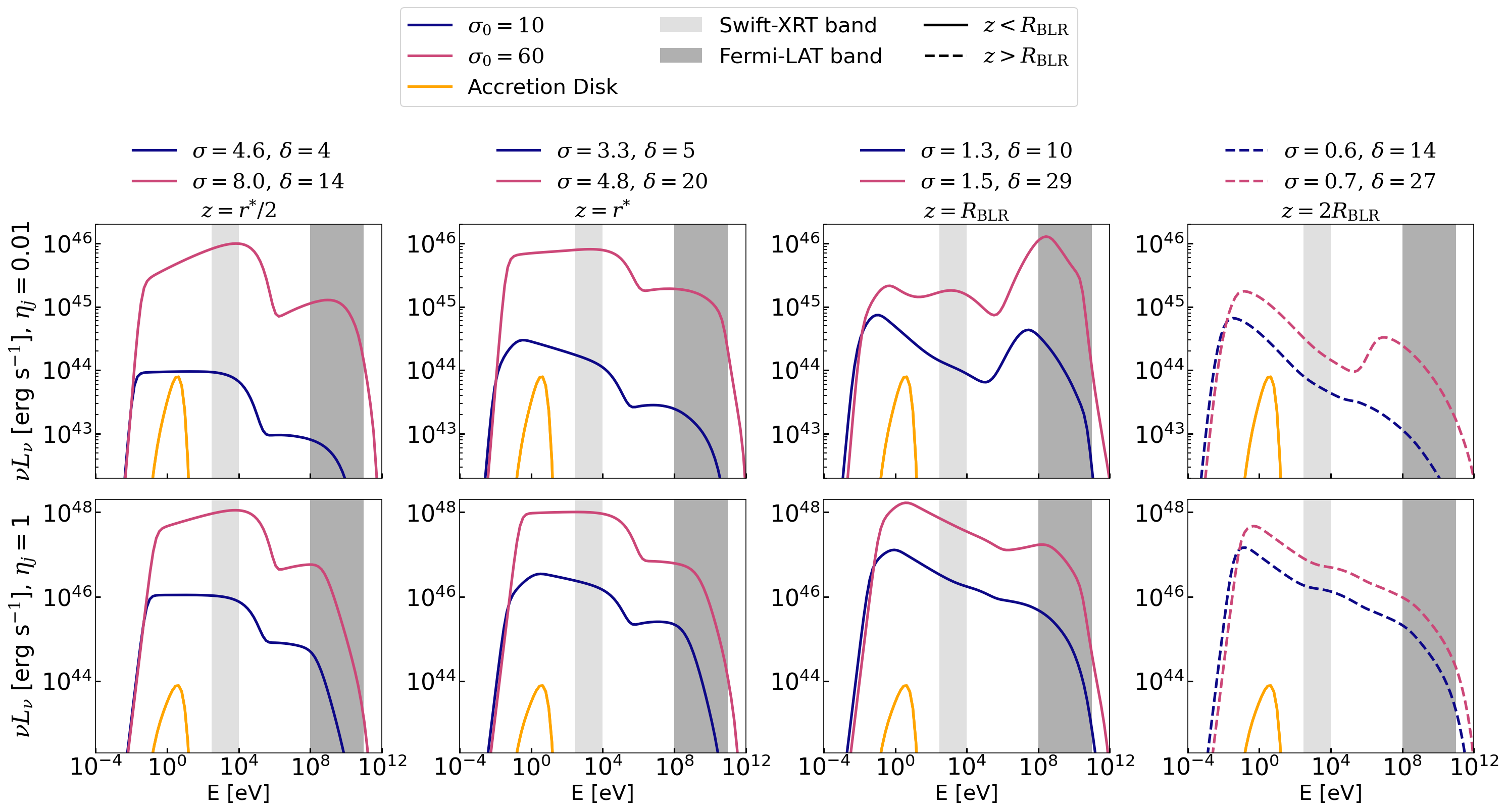} 
\caption{SEDs for different initial magnetizations $\sigma_0$, shown at four distances from the central SMBH (columns) and for two values of $\eta_j$ (rows). Columns (left to right) correspond to emission region locations at $z = r^*/2$, $z = r^*$, $z = R_{\rm BLR}$, and $z = 2, R_{\rm BLR}$. Solid curves denote emission inside the BLR ($z < R_{\rm{BLR}}$) while dashed curves denote outside ($z > R_{\rm{BLR}}$). In all panels, the multi-temperature accretion disc spectrum (as measured in the AGN frame) is also shown.}
\label{fig:SEDs_sigma_0}
\end{figure*}

\subsubsection{Variation of Initial Magnetization $\sigma_{0}$}\label{sec:sigma0}

We vary the initial magnetization $\sigma_0$ for two values, $10$ and $60$, while keeping other variables fixed to the baseline model. The initial magnetization sets the available energy content of the magnetic field to accelerate the bulk of the jet and the power-law index of the accelerated particle distribution. 

In Fig.~\ref{fig:SEDs_sigma_0} we present the SEDs for two values of initial magnetization $\sigma_0$ and $\eta_{\rm j}$. The SEDs are shown at four representative distances from the SMBH (columns): $z = r^*/2$, $z = r^*$, $z = R_{\rm BLR}$, and $z = 2 R_{\rm BLR}$, where $r^*$ is defined by Eq.~\ref{Eq:z_BLR_AD}. In each panel, colours indicate initial $\sigma_0$ values, and the line style (solid or dashed) denotes whether the emission region is located inside ($z < R_{\rm BLR}$) or outside ($z > R_{\rm BLR}$) the BLR. For reference, we include the observable energy ranges of Swift-XRT (X-ray) and Fermi-LAT ($\gamma$-ray), marked by shaded bands in each plot. Across each row of Fig.~\ref{fig:SEDs_sigma_0}, we observe how the spectral hardness and the peak energies of the SED components vary with the distance of the emitting region from the SMBH, similar to what we observe for the baseline model in Fig.~\ref{fig:baseline_vs_z}.  

The initial magnetization $\sigma_0$ strongly affects the spectral shape and the apparent brightness of the SED in three main ways: hardness of the particle distribution, power channelled into accelerated particles, and bulk acceleration of the jet. At the same distance $z$, jets with higher initial magnetization will also have higher $\sigma$, leading to a harder particle spectrum as described in Sec.~\ref{Sec:Particle_distribution}. 

Additionally, jets with higher $\sigma_0$ achieve larger terminal Lorentz factors $\Gamma_{\rm max}$. As a result, both $\Gamma$ and $\delta$ are expected to be larger at a given distance $z$ for jets with higher $\sigma_0$. The larger Doppler factor has two observable impacts on the SED: it shifts the spectral peaks to higher energies and increases the observed luminosity ($\nu L_\nu \propto \delta^4$). This is evident in Fig.~\ref{fig:SEDs_sigma_0} by comparing the highest $\sigma_0$ curve with the lowest $\sigma_0$ curve in the same panel; the high-$\sigma$ SED is typically brighter and harder (especially for distances $z<R_{\rm BLR}$). Higher $\sigma_0$ also leads to higher boosting of the external photon fields and thus to a more prominent IC component at the same distance from the SMBH. 

Regarding the effect of $\eta_{\rm j}$, for the case of $\eta_{\rm j}=0.01$ row (top), we find that the Compton dominance (ratio of IC peak to synchrotron peak) is much higher than in the $\eta_{\rm j}=1$ row, which is a direct consequence of weaker magnetic fields and thus lower synchrotron emission. Since $\eta_{\rm j}$ refers to the ratio of jet power to accretion power, higher $\eta_{\rm j}$ leads to higher Poynting fluxes of the jet and thus higher injected particle luminosities in the emitting region. If the charged particles cool and emit their energy as radiation, we expect higher bolometric luminosities for higher $\eta_{\rm j}$, something that is observed in Fig.~\ref{fig:SEDs_sigma_0}.

\subsubsection{Variation of Acceleration Efficiency $\eta_{\rm acc}$}\label{sec:eta_acc}

\begin{figure*}
\centering
\includegraphics[width=0.95\textwidth]{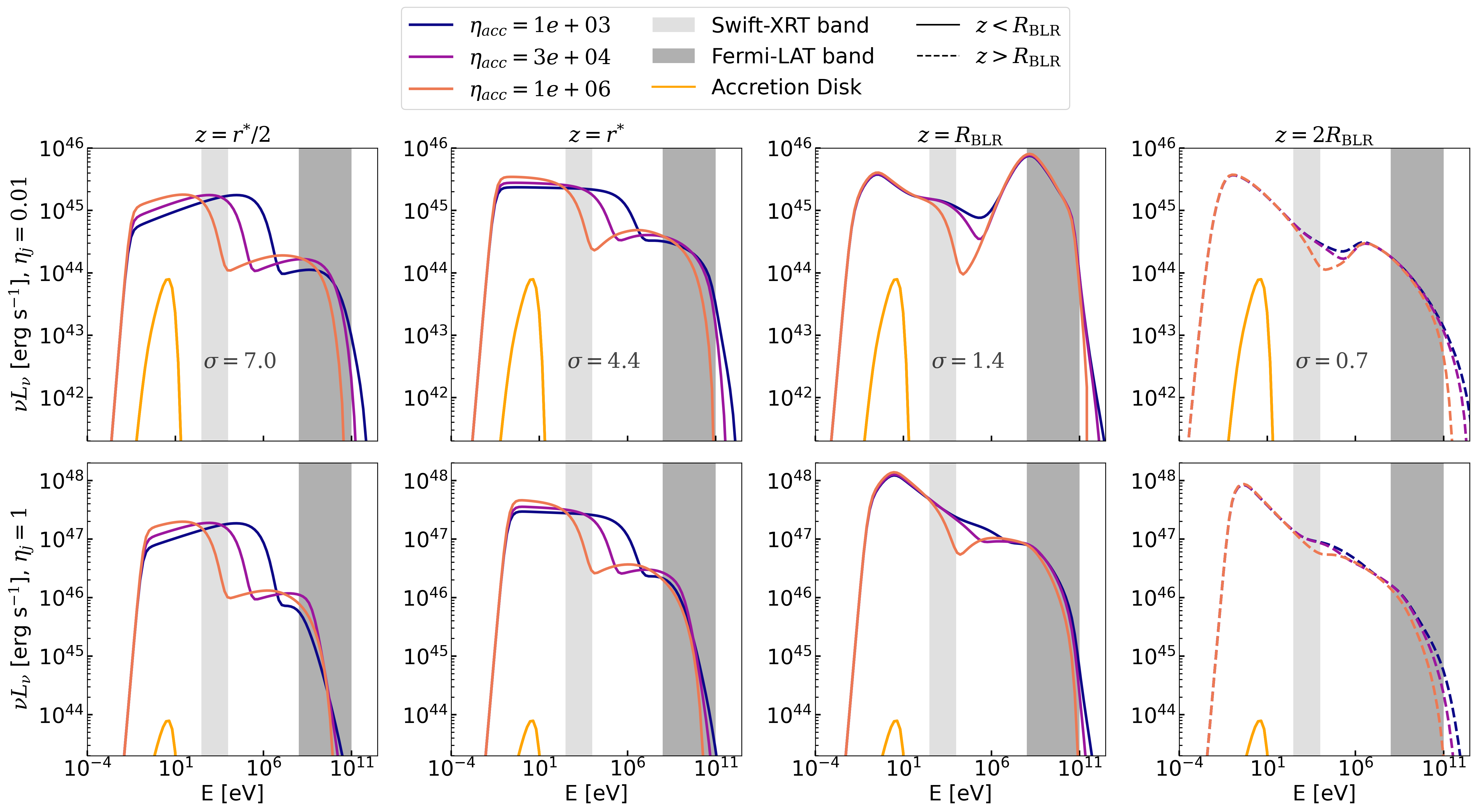} 
\caption{SEDs for different acceleration efficiencies $\eta_{\rm acc}$, shown at four distances from the central SMBH (columns) and for two values of $\eta_j$ (rows). Columns (left to right) correspond to emission region locations at $z = r^*/2$, $z = r^*$, $z = R_{\rm BLR}$, and $z = 2 R_{\rm BLR}$. Solid curves denote emission inside the BLR ($z < R_{\rm{BLR}}$) while dashed curves denote outside ($z > R_{\rm{BLR}}$). In all panels, the multi-temperature accretion disc spectrum (as measured in the AGN frame) is also shown.}
\label{fig:SEDs_eta_acc}
\end{figure*}

In Fig.~\ref{fig:SEDs_eta_acc} we present the SEDs for different particle acceleration efficiencies $\eta_{\rm acc}$\footnote{The adopted high value of $\eta_{\rm acc}=10^6$, corresponding to very slow particle acceleration, is necessary to shift the synchrotron peak below the keV energy band and thus to reproduce intermediate and low-synchrotron-peaked blazars, as illustrated in Fig.~\ref{fig:SEDs_eta_acc} (first column), for high magnetizations (hard pair distributions). However, this value is significantly larger (implying slower acceleration) than the acceleration efficiency ($\eta_{\rm acc}\sim10$) typically found in PIC simulations of magnetic reconnection \citep[e.g.,][]{2014ApJ...783L..21S}. Such efficient acceleration predicted by PIC simulations naturally leads to synchrotron peaks near the burnoff limit ($\sim$100 MeV), set by radiation-reaction losses, corresponding to extremely high synchrotron-peaked blazars.} at several distances from the SMBH, and for two values of $\eta_{\rm j}$ ($\eta_{\rm j}=0.01$ in the top row and $\eta_{\rm j}=1$ in the bottom row). In particular, larger $\eta_{\rm acc}$ results in a longer particle acceleration timescale (Eq.~\ref{Eq:t_acc}), thereby limiting the maximum electron Lorentz factor to lower values. This, in turn, shifts the synchrotron peak to lower energies from the MeV to the keV energy band. We find the same behaviour for the high-energy component of the SED with increasing $\eta_{\rm acc}$ for distances $z=r^{*}/2$ (first column) and for $\eta_{\rm j}=0.01$(first row) as the average energy of the synchrotron photon distribution is smaller, thus the up-scattered photons will have on average lower energy too. This feature is also present at a distance z=$r^{*}$ from the SMBH, but due to $\gamma \gamma$ absorption, it is not directly visible (for $\eta_{\rm j}=1$). However, a small change can be observed near the cutoff at about 100 GeV. Comparing the two $\eta_{\rm j}$ cases, we find that the high efficiency jet ($\eta_{\rm j}=1$, bottom row) produces a more luminous SED overall than the low efficiency jet ($\eta_j=0.01$, top row), but both follow the same qualitative trend with respect to $\eta_{\rm acc}$. In the $\eta_{\rm j}=1$ case, the synchrotron and IC peaks remain more luminous for a given $\eta_{\rm acc}$ (due to the greater energy budget in the jet), but we find a shift to lower energies as $\eta_{\rm acc}$ increases. For the case of z=$R_{\rm BLR}$ (third column in~Fig.\ref{fig:SEDs_eta_acc}) where the external field of the BLR is responsible for the EC emission we find that different $\eta_{\rm acc}$ give the same high-energy emission component.

\subsubsection{Variation of Mass Accretion Rate $\dot{m}$}\label{sec:mdot}
\begin{figure*}
\centering
\includegraphics[width=0.95\textwidth]{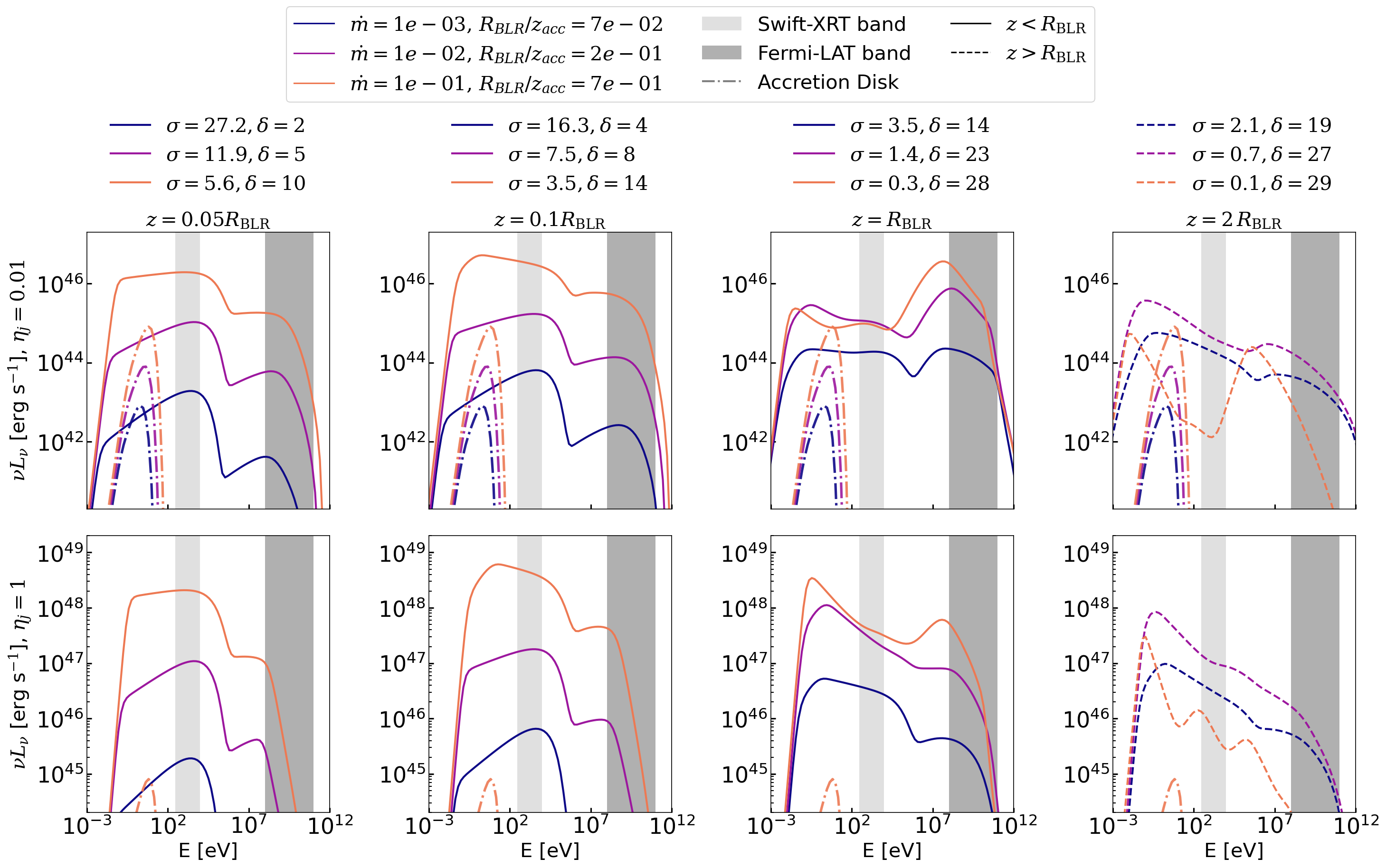} 
\caption{SEDs for different mass accretion rates $\dot{m}$, shown at four distances from the central SMBH (columns) and for two values of $\eta_j$ (rows). Columns (left to right) correspond to emission region locations at $z = 0.05R_{\rm BLR}$, $z = 0.1R_{\rm BLR}$, $z = R_{\rm BLR}$, and $z = 2, R_{\rm BLR}$. Solid curves denote emission inside the BLR ($z < R_{\rm{BLR}}$) while dashed curves denote outside ($z > R_{\rm{BLR}}$). Dash-dotted lines in each panel correspond to the AD spectrum in the AGN rest system.}
\label{fig:SEDs_dot_m}
\end{figure*}

To examine how the mass accretion rate $\dot{m}$ affects the SEDs, we present in Fig.~\ref{fig:SEDs_dot_m} simulations performed at fixed relative distances to the BLR (columns), i.e., constant $z/R_{\rm BLR}$, for different values of $\dot{m}$. Because the BLR radius changes with the AD luminosity and thus the mass accretion rate as $R_{\rm BLR} \propto \dot{m}^{1/2}$, by altering the latter we sample emitting regions at different absolute distances $z$ from the SMBH. Specifically, lower accretion rates correspond to physically smaller BLRs, placing the emitting region closer to the jet base, which is characterized by higher magnetization $\sigma$ and thus harder pair injection spectra. Conversely, higher $\dot{m}$ places the emitting region farther out, into environments of lower magnetization and softer particle injection spectra for SEDs in the same panel. For these simulations, we account for all radiative processes. 

The differences in $\sigma$ and $\delta$ arising from the location change, together with the changing external photon density, translate directly into differences in the model SEDs for each $\dot{m}$. At small relative distances (left columns), the emitting region is located close to the SMBH and within the characteristic acceleration zone ($z < z_{\rm acc}$) for all cases, thus achieving only modest bulk Lorentz factors $\Gamma$ due to incomplete magnetic to kinetic energy conversion. In these regions, despite the high photon density of the AD and the BLR, the external photon fields are not strongly boosted, limiting the effectiveness of the EC process. Consequently, synchrotron emission and SSC dominate, while the EC component remains comparatively moderate. As the emission region moves outward, closer to $z_{\rm acc}$, the maximal bulk Lorentz factor is attained, strongly boosting the external photon (mainly BLR contribution) fields and significantly enhancing the EC emission (middle columns). In these intermediate distances (closer to the BLR boundary, i.e., $z\approx R_{\rm BLR}$), EC emission becomes most pronounced due to the combination of high bulk Lorentz factors and dense photon fields from the BLR. Consequently, the SED becomes EC-dominated, exhibiting bright GeV emission and a prominent Compton peak. At even larger distances (rightmost columns), beyond the BLR ( $z>R_{\rm BLR}$ ), the external photon fields are weaker since they are deboosted. As a result, the emission is dominated by synchrotron, SSC processes, and some EC contributions from the presence of the IR DT in the MeV band. 
  
\subsection{Leptohadronic model} \label{sec:leptohadronic}

Next we explore the impact of including protons in the emitting region and how this affects the SED. Because the power–law index of charged particles accelerated via magnetic reconnection depends on the local magnetisation, $\sigma(z)$, the injected proton energy distribution becomes softer as the distance from the SMBH increases.  
When $\sigma \lesssim 3$, most of the proton energy goes in low–$\gamma$ particles $\gamma_{\rm p}\simeq\gamma_{\rm p,min}$, whereas for $\sigma \gtrsim 3$ the energy is concentrated near the maximum Lorentz factor, $\gamma_{\rm p,\max}$.  
To illustrate these two regimes, we place the emitting region at two characteristic distances using baseline parameters: close to the BLR at $0.1R_{\rm BLR}$ (high‐$\sigma\simeq8$) and farther out at $R_{\rm BLR}$ (low‐$\sigma\simeq1$).

\begin{figure*} 
\includegraphics[width = 0.49\textwidth]{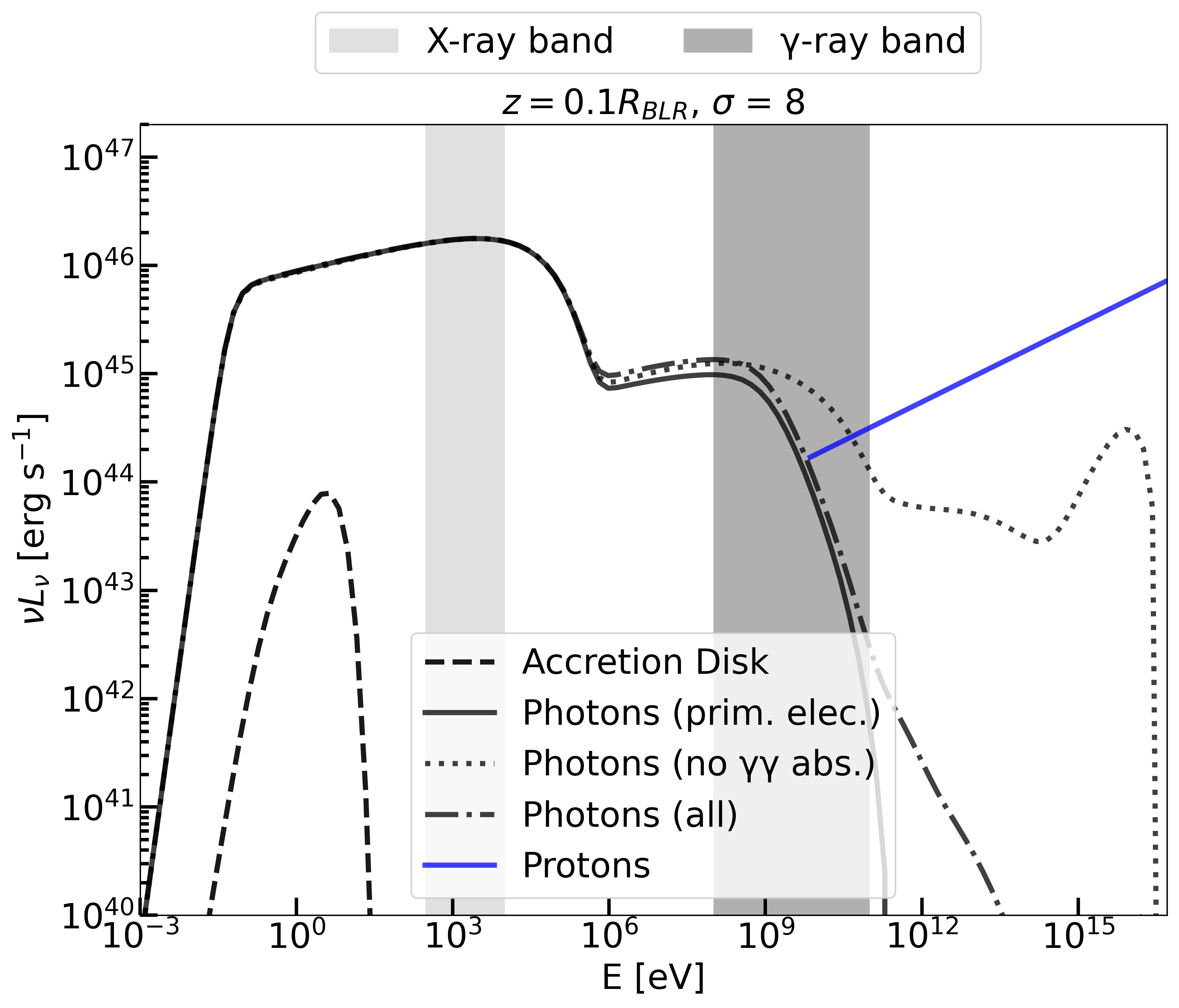}
\includegraphics[width = 0.49\textwidth]{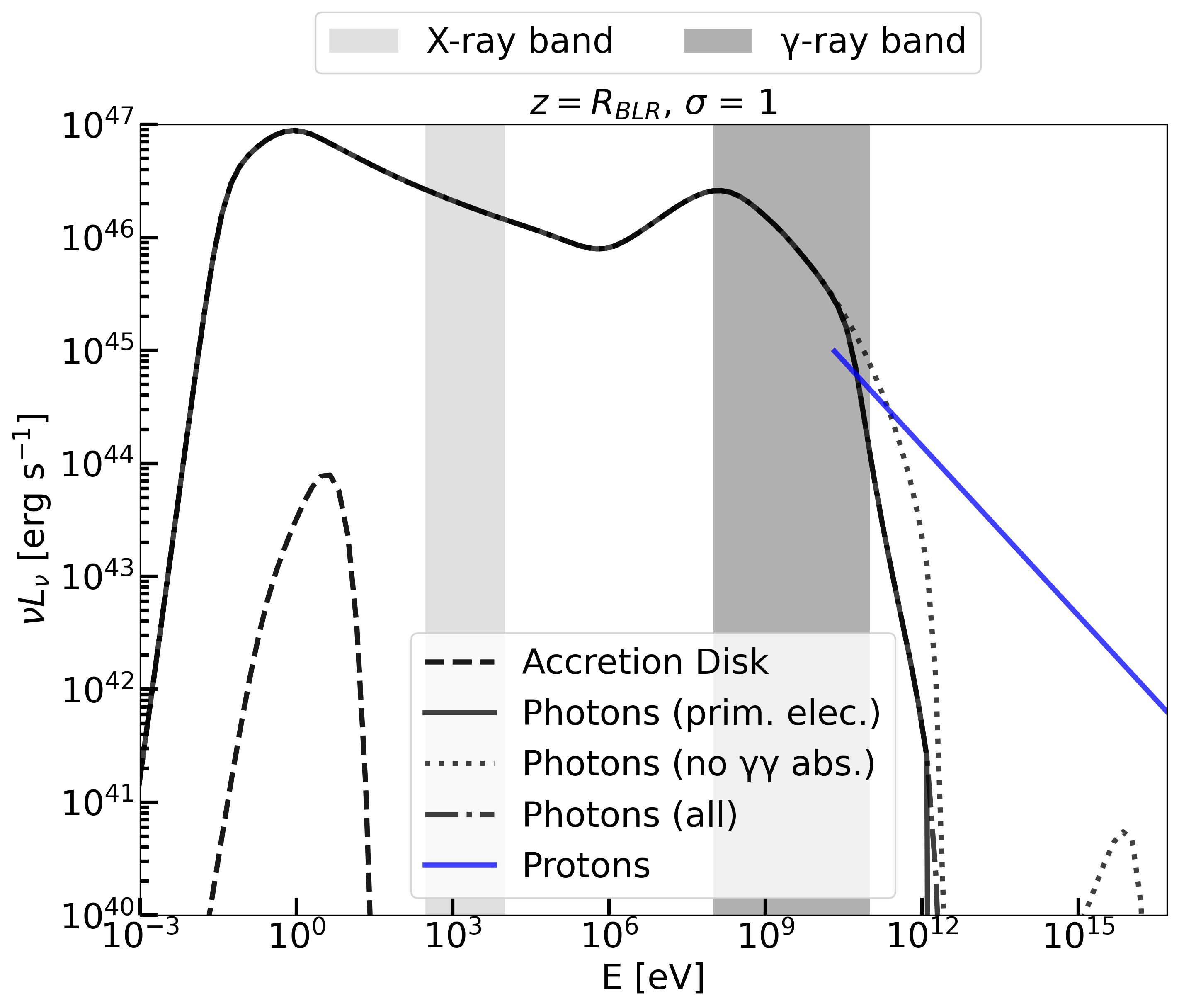}
\caption{Broadband SEDs with and without protons. Solid curves show the purely leptonic solution, while dash–dotted curves include a co-spatial proton population. Dotted curves repeat the leptohadronic spectra with internal $\gamma\gamma$ absorption switched off. Shaded vertical bands mark the Swift–XRT (light grey) and Fermi–LAT (dark grey) observing windows. The SED on the left panel is calculated for the baseline values of the parameters at distance $z=R_{\rm BLR}$ (low-$\sigma$ case) while the SED on the right panel is calculated at distance $z=0.1 R_{\rm BLR}$ (high-$\sigma$ case)}
\label{fig:sigma_lh_comp}
\end{figure*}

Fig.~\ref{fig:sigma_lh_comp} summarizes the results. In the low‐$\sigma$ case (right panel) the leptohadronic spectrum (dash–dotted) is identical to the purely leptonic one (solid), while the emission from protons (proton synchrotron, secondary $e^\pm$ emission, and $\gamma$–rays from neutral–pion decay) falls below the dominant SSC emission of the primary pairs. Only when internal $\gamma\gamma$ absorption is artificially turned off (dotted line) does the $\pi^0\ \rightarrow\ 2\gamma$ component emerge at $\sim10^{16}\,$eV. These photons are attenuated in-source by lower-energy photons. The luminosity of the $\pi^0$ component (peaking at $\sim 10$ PeV) in this case is driven by the shape of the proton distribution. In the low-magnetization regime, protons attain a soft spectrum, leading to inefficient pion production and thus suppressed $\pi^0 \to \gamma$ emission.

In the high‐$\sigma$ case (left panel) exhibits noticeable differences between leptonic and leptohadronic scenarios, particularly in the MeV to GeV range. The emission at these energies is attributed to proton synchrotron radiation and IC Compton scattering of low-energy photons from pairs (the same has already been discussed in \cite{Petro2023}). Beyond $\sim 10$~GeV, the spectrum from secondary $e^\pm$ and $\pi^0$ decay is again suppressed by $\gamma\gamma$ absorption (compare dash–dotted and dotted curves).

\section{Neutrino production for the baseline model}\label{sec:neutrino_prod}
\begin{figure}
\centering
\includegraphics[width=0.45\textwidth]{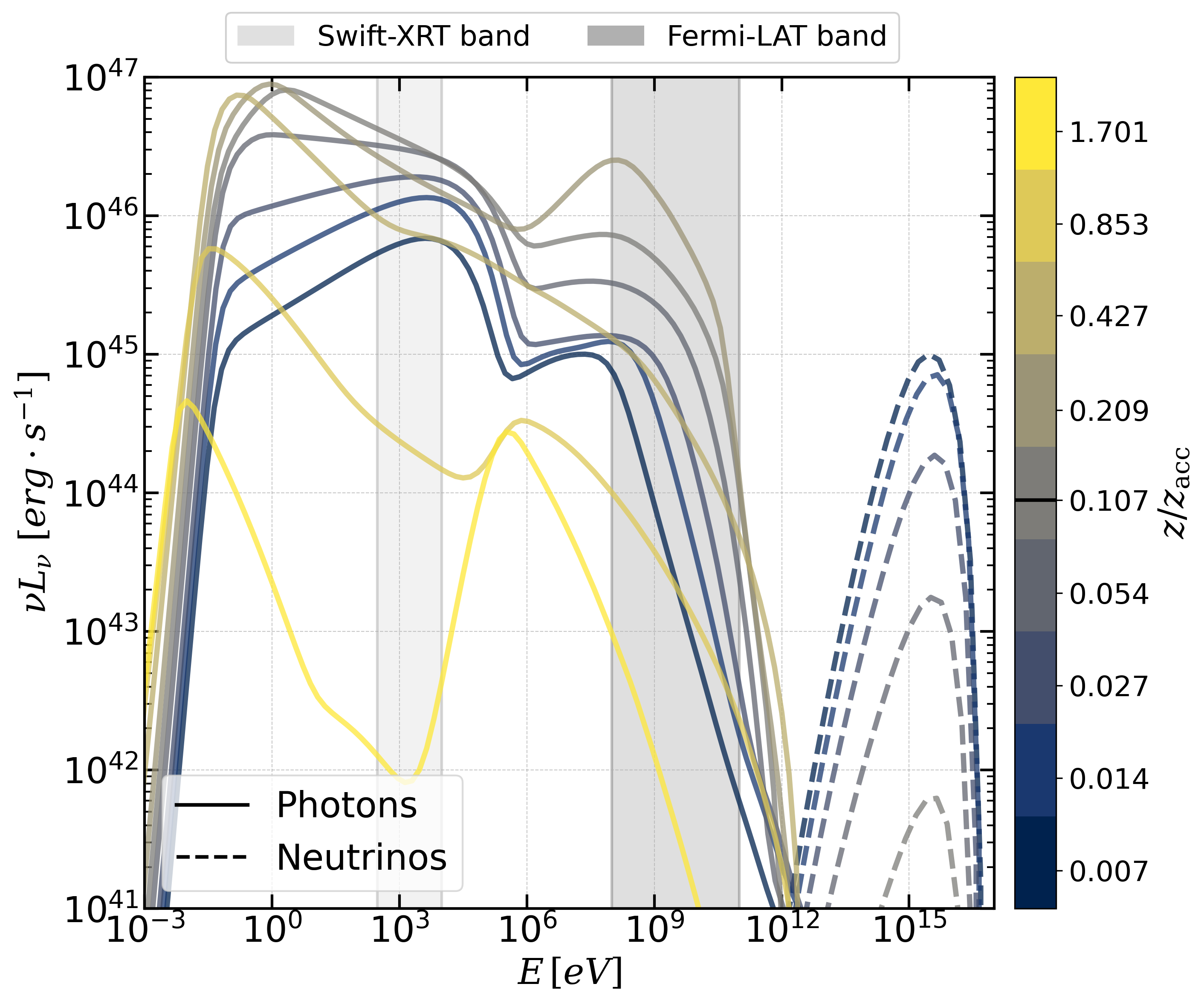} 
\caption{SEDs of photons (solid lines) and neutrinos (dashed lines) for the baseline model as a function of distance from the SMBH, same as Fig.\ref{fig:baseline_vs_z}. The shaded light gray and darker gray bands highlight the 0.3–10~keV and 100 MeV–100 GeV energy ranges, respectively.}
\label{fig:baseline_nu_vs_z}
\end{figure}

Next, we explore neutrino production in our model. In Fig.~\ref{fig:baseline_nu_vs_z}, we present the evolution of the SEDs for the baseline model, including neutrino emission (dashed lines) as a function of distance from the SMBH. The emission is boosted to the observer according to the Lorentz factor of the location of the unperturbed flow, and different colors indicate the normalized jet distance $z/R_{\rm BLR}$ as in Fig.~\ref{fig:baseline_vs_z}. 

\begin{figure}
\centering
\includegraphics[width=0.45\textwidth]{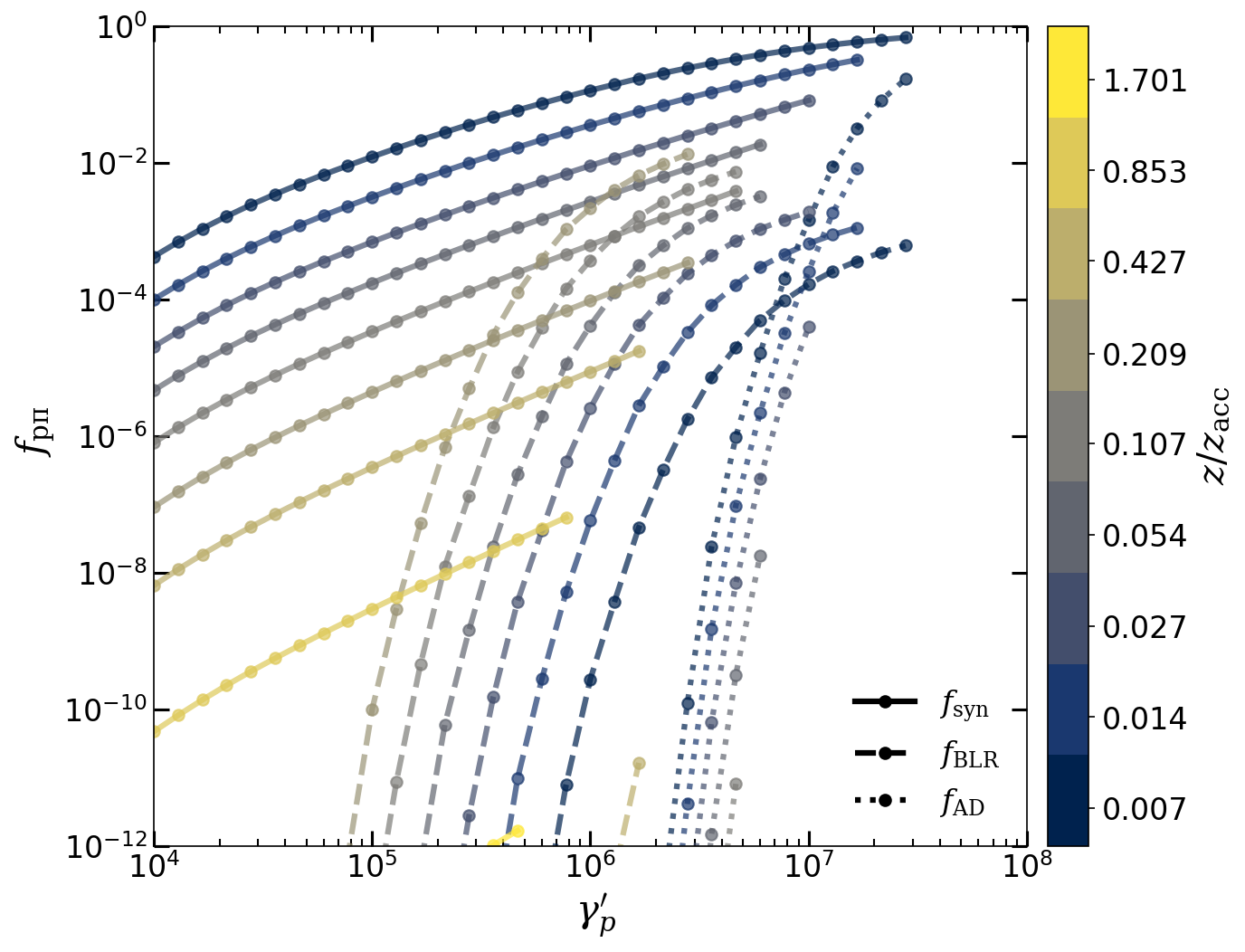} 
\caption{Photopion production efficiency $f_{\rm p \pi}$ as a function of proton Lorentz factor $\gamma_{\rm p}$ and for various distances $z$ from the SMBH. Different linestyles are used to indicate different pion production efficiencies for three target photon fields: jet synchrotron (solid lines), BLR (dashed lines), and AD radiation (dotted lines).}
\label{fig:f_pg}
\end{figure}

The neutrino emission for our baseline model peaks at an observed energy, 
\begin{equation}
E_{\nu}(z) \approx \frac{1}{20}\,\delta(z)\,\gamma'_{\rm p,max}(z)\,m_{\rm p}c^2 \sim 4\,\mathrm{PeV},
\label{Eq:Observed_nu_energy}
\end{equation}
as the power-law index of the proton distribution is $p<2$. The black line in the colourbar of Fig.~\ref{fig:baseline_nu_vs_z} denotes the distance at which the power-law index of the particle distributions in the emitting region becomes $2$. For smaller distances, the power-law index becomes smaller (harder spectrum). We find that the observed neutrino luminosity remains approximately constant as long as the emitting region lies upstream of the BLR and the power-law index of the proton distribution is $p<2$. In this regime, photons inside the emitting region, which consist of the synchrotron, the BLR, and AD (Doppler boosted into the jet's comoving frame), are the main targets for photopion interactions. However, once the emitting region is located beyond the BLR, the contribution of external photons becomes negligible due to relativistic deboosting. In this case, the dominant target photon field becomes the synchrotron radiation produced internally by mostly the primary pairs, which weakens with distance because the magnetic field strength decreases along the jet. The observed all-flavor neutrino luminosity can be approximated as follows, 
\begin{equation}
E_{\nu}L_{\nu+\bar{\nu}}(E_{\nu}) \approx \frac{3}{8} f_{\rm p\pi}(E_{\rm p}) L_{\rm p}(E_{\rm p})=Y_{\nu \gamma}L_{\gamma},
\label{Eq:Observed_lum}
\end{equation}
where we assume that we assume that the protons energy is connected with the neutrino energy as  $E_{\rm p} \sim 20 E_{\rm \nu}$, and $f_{\rm p \pi} \equiv \left(1 + \frac{t'_{\rm p \pi}}{t'_{\rm cr}}\right)^{-1}$ is the photopion production efficiency, $t'_{\rm p \pi}$ is the proton energy loss timescale due to photomeson interactions, and $t'_{\rm cr}$ is the characteristic escape timescale, which we approximate as the light-crossing time of the region, and $Y_{\nu \gamma}\equiv L_{\nu+\bar{\nu}}/L_{\gamma}$. In Fig.~\ref{fig:f_pg} we present $f_{\rm p \pi}$ as a function of the proton Lorentz factor $\gamma_{\rm p}$ for various emitting region locations. Different linestyles correspond to the pion production efficiency for three target photon fields, jet synchrotron (solid lines), BLR (dashed lines), and AD radiation (dotted lines). We find that for our baseline model, the most dominant targets for photon-pion interactions are synchrotron photons. 

\begin{figure}
\centering
\includegraphics[width=0.45\textwidth]{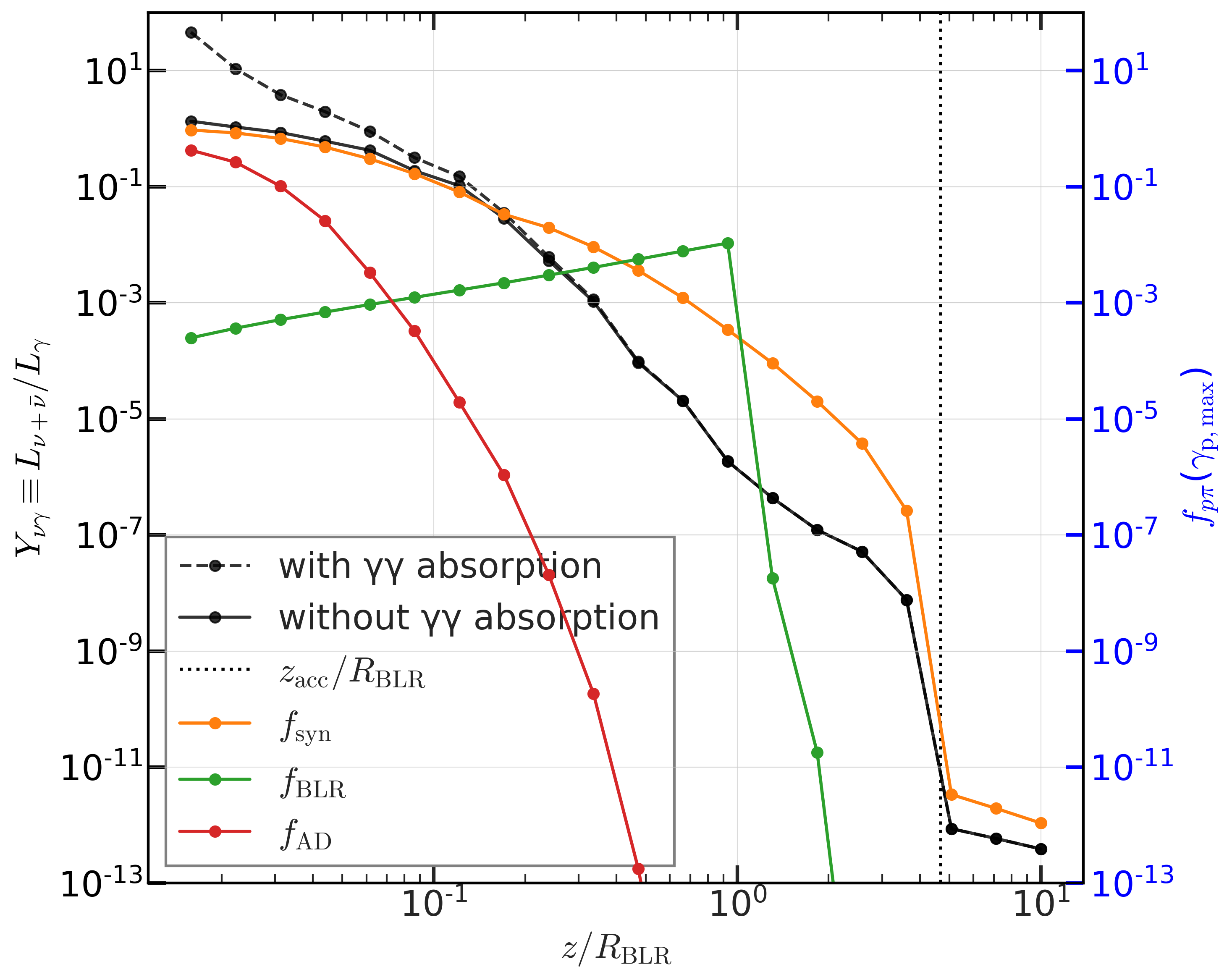} 
\caption{Black lines (left ordinate) give the neutrino to $\gamma$-ray luminosity ratio, $Y_{\nu\gamma}$, calculated with (dashed) and without (solid) internal $\gamma\gamma$ absorption. Coloured curves (right ordinate) show the pion production efficiency for the maximum proton energy of the proton distribution for three target photon fields: jet synchrotron (orange $f_{\mathrm{syn}}$), BLR (green $f_{\mathrm{BLR}}$), and AD radiation (red $f_{\mathrm{AD}}$). The vertical dotted line marks the end of the magnetic acceleration zone at $z=z_{\mathrm{acc}}$. }
\label{fig:ratio_lum_f_pg}
\end{figure}

On the left y-axis of Fig.~\ref{fig:ratio_lum_f_pg} we present the ratio $Y_{\nu \gamma}$ with and without $\gamma \gamma$ absorption (dashed and solid black curves, respectively). We find that for distances $z<0.1R_{\rm BLR}$ the neutrino luminosity overshoots the $\gamma$-ray luminosity up to a factor $\sim 10$ since the latter is internally absorbed by lower energy photons. The ratio $Y_{\nu \gamma}$ is a decreasing function of distance since the neutrino emission tracks the proton distribution inside the emitting region (see Eq.~\ref{Eq:Observed_lum}) and the $\gamma$-ray luminosity, which, depending on the distance, is attributed to SSC, proton synchrotron and EC. 

The proton luminosity in the observer's frame is given by $L_{\rm p} = \delta^4 L'_{\rm p}$. On the right y-axis of Fig.~\ref{fig:ratio_lum_f_pg} we present the pion production efficiencies $ f_{\rm p\pi}$ for proton interactions with: (i) the co-moving synchrotron photon field (orange), (ii) the BLR field (green), and (iii) the AD field (blue). These efficiencies are evaluated at the maximum proton Lorentz factor $\gamma_{\rm p,max}$, which is responsible for the highest neutrino production, obtained from the balance between acceleration and cooling (see Sec.~\ref{Subsec:Radiative_losses}). We find that there is no single target photon field that dominates everywhere. Specifically, the AD photons control the pion production for the first few Schwarzschild radii, synchrotron photons for the bulk of the acceleration zone, and BLR photons only for a narrow range of distances around 0.3-1 $R_{\rm BLR}$. In Appendix~\ref{app:nu_on_BLR} we present analytical considerations on the neutrino production, assuming that the target photon fields consist of monochromatic BLR photons. 

\section{Model Predictions and Blazar Observations}\label{sec:observations}
\begin{figure}
\centering
\includegraphics[width=0.45\textwidth]{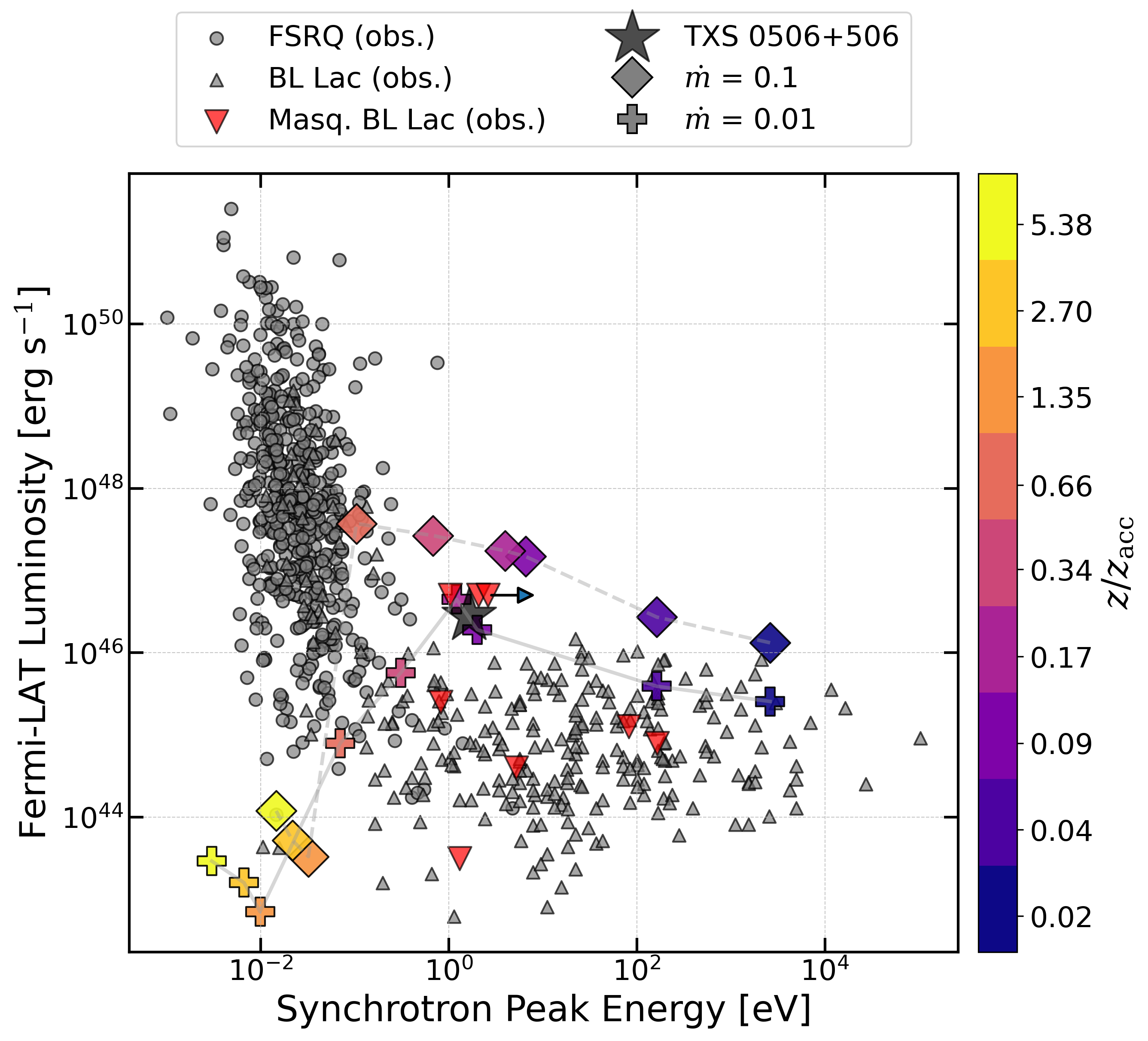} 
\caption{Comparison between observed and model-predicted theoretical gamma-ray luminosities as a function of the synchrotron peak energy for the baseline model. Gray points represent observational data from~\citep{2023ApJ...944..157C}: FSRQ are shown as circles, and BL Lac objects as triangles.  Red triangles
represent masquerading BL Lac sources (adopted from \citep{2022MNRAS.511.4697P}). The average value for TXS 0506+056 is adopted from \citep{2019MNRAS.484L.104P}. The theoretical model predictions are calculated at the same distances from the SMBH as in Fig.~\ref{fig:baseline_nu_vs_z} and are shown with different colors. Lines that connect the points illustrate the progression of synchrotron peak energy and gamma-ray luminosity.}
\label{fig:L_vs_peak_baseline}
\end{figure}

We further investigate how our model predictions compare with observations of Fermi-detected blazars in Fig.~\ref{fig:L_vs_peak_baseline}. The sample used for comparison consists of 781 Fermi-detected AGN, which includes 504 FSRQs and 277 BL Lacs
objects~\citep{2023ApJ...944..157C} and is represented as gray points in Fig.~\ref{fig:L_vs_peak_baseline} (circles represent FSRQs and triangles represent BL Lac objects). Red triangles represent masquerading BL Lac sources, while we also include the average values of the synchrotron peak and $\gamma-$ray luminosity detected in the Fermi-LAT band for the TXS 0506+056 as adopted from~\cite{2019MNRAS.484L.104P} (see black star in Fig.~\ref{fig:L_vs_peak_baseline}).

Coloured markers indicate model expectations based on the location of the emitting region along the jet. All model parameters have their baseline values except for the mass accretion rate $\dot{m}$, for which we considered two indicative values. The location of the emitting region affects the synchrotron peak energy and the $\gamma$-ray luminosity of the SED. In particular, the synchrotron peak shifts from keV energies, typical of low-luminosity HSP BL Lacs, toward lower energies. Initially, the luminosity in the Fermi band increases due to reduced internal $\gamma\gamma$ absorption, a decreasing density of external photon fields (both synchrotron and AD photons, which serve as targets for Fermi-band $\gamma$-rays), and increasing Doppler boosting ($\delta$). Once the particle power-law index crosses $p<2$ (closer to the SMBH), synchrotron peaks shift to IR/optical (characteristic of LSP BL Lacs), and Fermi luminosity peaks near the BLR. Outside the BLR, $ \gamma$-ray luminosity in the Fermi-LAT becomes negligible, since the EC emission peaks in the $\sim$ MeV energy band. Overall, the baseline model successfully reproduces the majority of BL Lacs, but fails to explain the high-luminosity LSP/FSRQs. Increasing the mass accretion rate $\dot{m}$ enhances the external photon field luminosity (AD and BLR), amplifying the EC seed photon density. Consequently, $\gamma$-ray luminosity in the Fermi-LAT band increases with $\dot{m}$. Thus, we find that by increasing $\dot{m}$ the model begin to populate the high $L_{\rm \gamma}$, low-synchrotron peak (LSP/FSRQ) part of the diagram, though full coverage of the FSRQ population still requires additional model adaptations (e.g. adjustments to mass of the SMBH, $\sigma_0$, or jet efficiency).

\begin{figure}
\centering
\includegraphics[width=0.45\textwidth]{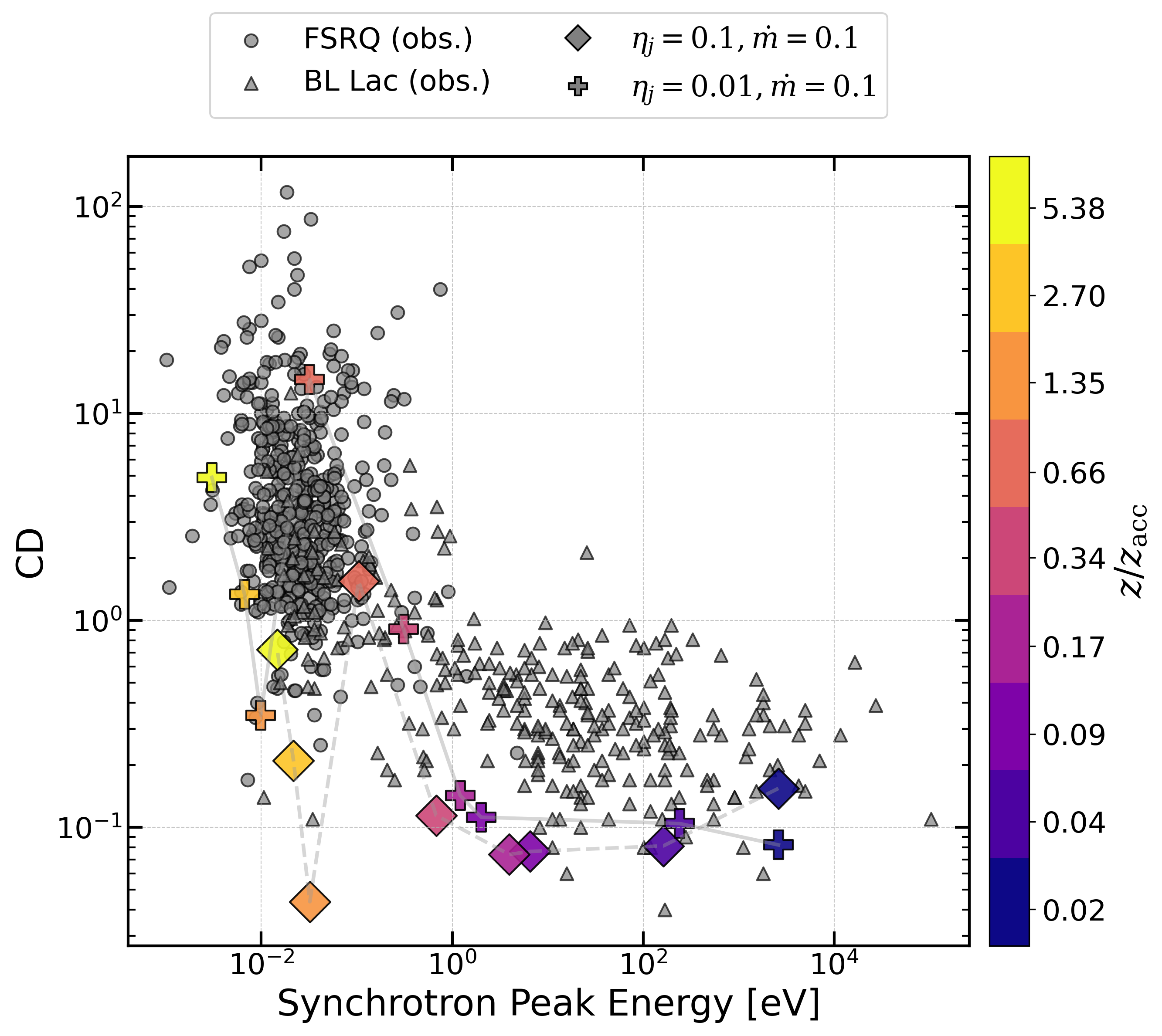} 
\caption{Comparison between observed and model-predicted theoretical Compton dominance (CD) as a function of the synchrotron peak energy for the baseline model. Gray points represent observational data from~\citep{2023ApJ...944..157C}: FSRQ are shown as circles, and BL Lac objects as triangles. The theoretical model predictions are calculated at the same distances from the SMBH as in Fig.~\ref{fig:baseline_nu_vs_z} and are shown with different colors. The dashed line illustrates the progression of synchrotron peak energy and gamma-ray luminosity.}
\label{fig:CD_vs_peak_baseline}
\end{figure}

Next, in Fig.~\ref{fig:CD_vs_peak_baseline}, we explore how Compton dominance (CD) varies when we modify the jet power efficiency $\eta_{\rm j}$, while fixing the mass accretion rate $\dot{m}=0.1$, and adopting the baseline values for the other parameters.
By reducing the value of $\eta_{\rm j}$ below its baseline value, we decrease the jet's magnetic field strength (see~\ref{Eq:B0}), which also leads to lower synchrotron luminosity and thus higher CD. Second, increasing $\dot{m}$ from the baseline value boosts the external photon field luminosity, which directly enhances EC contribution, further increasing the CD. For distances of the emitting region from the SMBH $z$ before the BLR, CD remains almost constant $\sim 10^{-1}$, varying within a factor of $~5$. We find that similar CDs are obtained from BL Lac objects in the sample of~\cite{2023ApJ...944..157C}. As the distance from the SMBH $z$ increases (decreasing magnetic field strength and stronger EC emission), we find that the model predictions shift upward and leftward in the CD versus synchrotron peak-energy plane, bringing them closer to the high-CD, low-synchrotron peak trends seen in luminous FSRQs. 

In the previous paragraph, we showed that the baseline model, which assumes a fixed SMBH mass and a fixed jet viewing angle, fails to reproduce the most luminous LSP/FSRQs. In this prescription, the highest $\gamma$-ray luminosities are obtained only when the emitting region lies close to the BLR, where the external photon density is stronger and IC scattering is more efficient when compared to the other jet's locations. A straightforward way to raise the BLR photon density, and therefore the $\gamma$-ray output, is to further increase the accretion rate, as demonstrated in Fig.~\ref{fig:L_vs_peak_baseline}. Doing so produces SEDs with $L_{\gamma}\gtrsim10^{47}\,\mathrm{erg\,s^{-1}}$ and synchrotron peaks at $\sim0.1\,\mathrm{eV}$.

\begin{figure}
\centering
\includegraphics[width=0.45\textwidth]{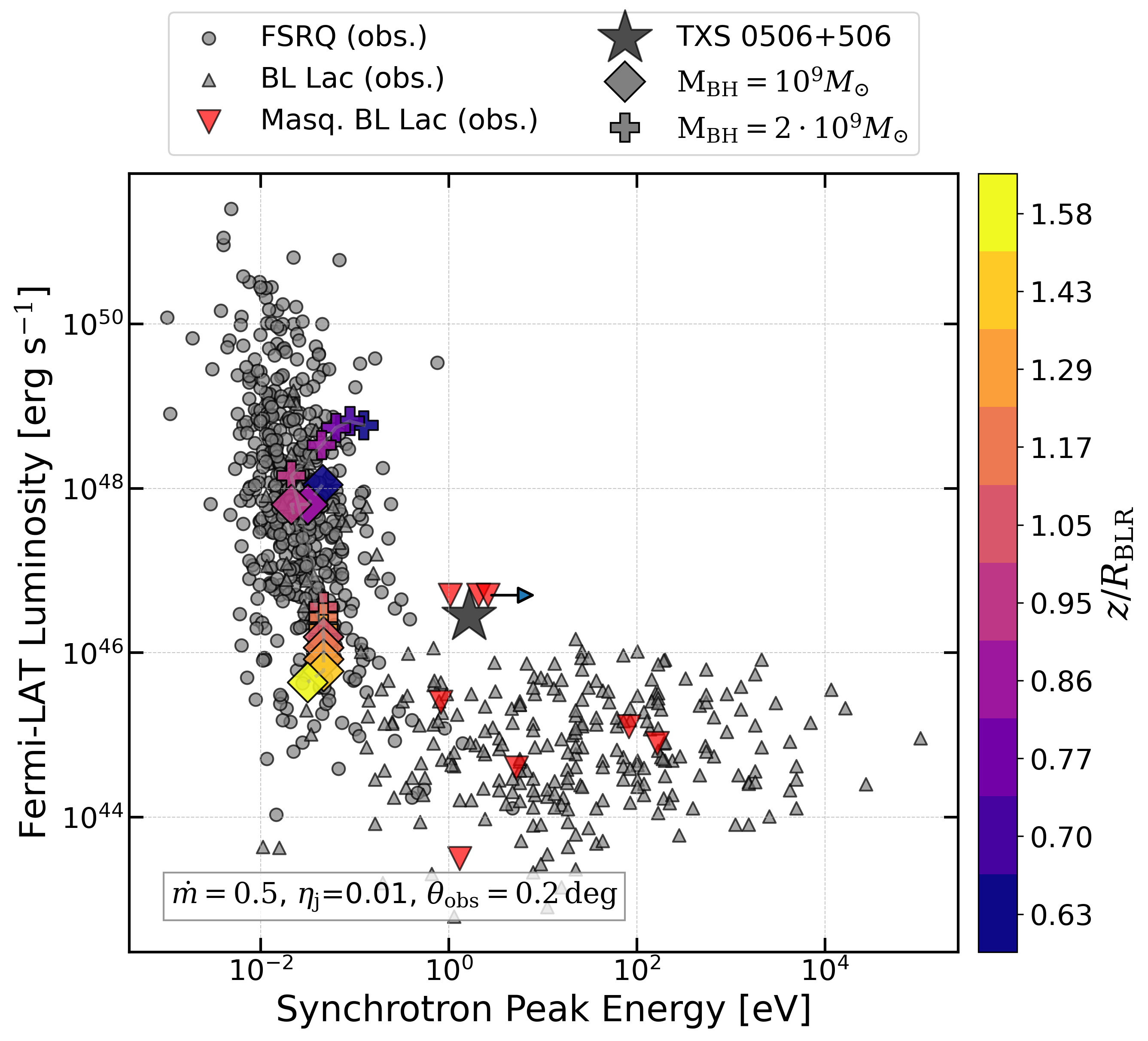} 
\caption{Same as in Fig.\ref{fig:L_vs_peak_baseline}. The theoretical model predictions are calculated for distances from the SMBH closer to the BLR and are shown with different colors. Lines that connect the points illustrate the progression of synchrotron peak energy and gamma-ray luminosity. All parameters have the baseline value except for the mass accretion rate, the viewing angle, and the mass of the SMBH. }
\label{fig:L_vs_peak_fsrq}
\end{figure}

The luminosity in the observer's frame can be boosted further by (i) reducing the viewing angle, thereby increasing the Doppler factor, and (ii) adopting a more massive SMBH together with a higher accretion rate, which amplifies all external radiation fields. Figure~\ref{fig:L_vs_peak_fsrq} illustrates such an alternative model realization, all parameters remain at their baseline values except for the accretion rate, the jet to accretion power ratio, and the viewing angle, while we show emitting regions that are close to the BLR. In this example, we decrease $\eta_{\rm j}$ from each baseline value since higher magnetic fields would lead to higher synchrotron peaks \footnote{The synchrotron peak energy is determined by synchrotron self absorption and it happens at the transition from the optically thick to the optically thin part of the spectrum. Higher magnetic field values at the given distance lead to higher synchrotron peak energies.}. Figure~\ref{fig:L_vs_peak_fsrq} confirms that a modest change in viewing angle ($0.2~\rm deg$) combined with a higher accretion rate ($\dot{m}=0.5$) is already sufficient to lift the model points into the observed FSRQ locus. Increasing $M_{\rm BH}$ from
$10^{9}\,M_{\odot}$ (diamonds) to $2\cdot 10^{9}\,M_{\odot}$ (pluses) moves the track further upward, allowing us to reproduce even brighter sources\footnote{Observationally, FSRQs objects typically have black-hole masses $\gtrsim7\times10^{8}\,M_{\odot}$ and this value can exceed a few $10^{9}\,M_{\odot}$ \citep{2013A&A...560A..28C}.} $L_{\gamma}\gtrsim10^{48}\,\mathrm{erg\,s^{-1}}$) without pushing the synchrotron peak above \(\sim10^{-2}\,\mathrm{eV}\). 

\begin{figure}
\centering
\includegraphics[width=0.45\textwidth]{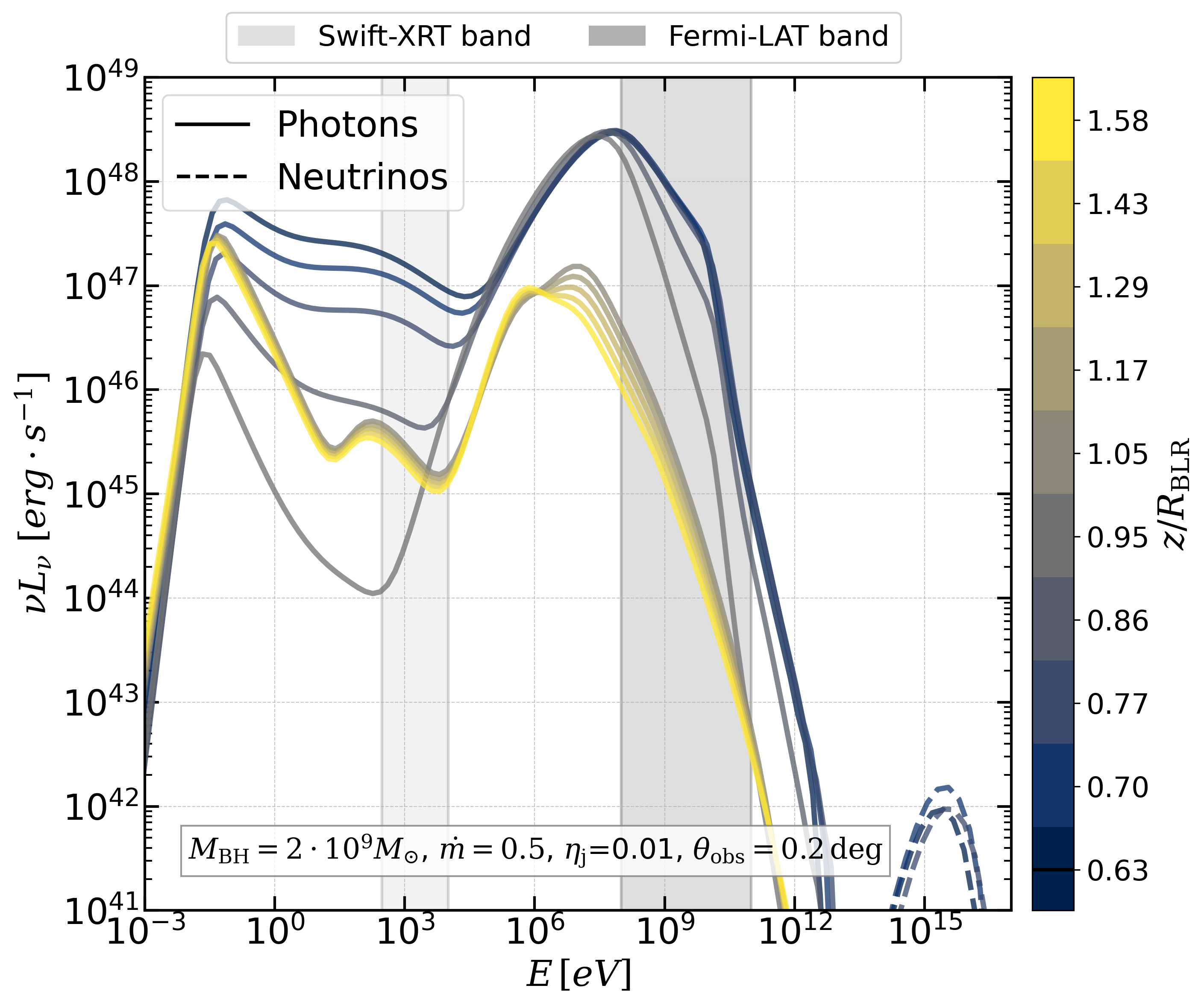} 
\caption{SEDs of photons (solid lines) and neutrinos (dashed lines) same as in Fig.~\ref{fig:baseline_nu_vs_z}. The parameters that are different compared to the baseline model are listed in the bottom legend.}
\label{fig:SEDs_fsrq}
\end{figure}

In Fig.~\ref{fig:SEDs_fsrq} we present the corresponding SEDs for the case with $M_{\rm BH}$ $2\cdot 10^{9}\,M_{\odot}$ (pluses in Fig.~\ref{fig:L_vs_peak_fsrq}). Since we are sampling locations in the jet that are close to the BLR region, we find enhanced EC emission while also a soft synchrotron spectrum with the peak of the synchrotron emission at $\sim 10^{-1}$~eV, as also shown in Fig.~\ref{fig:L_vs_peak_fsrq}. The spectrum in the Fermi-LAT band for the first three locations is a power law, whereas the feature found in almost all SEDs at $\sim$keV energies is due to SSC emission. The neutrino luminosity is relevantly low compared with the $\gamma$-ray luminosity observed in the Fermi-LAT band since the protons do not have a hard spectrum at these distances, as already mentioned in Sec.~\ref{sec:neutrino_prod}.

\section{Discussion}\label{sec:discussion}

In this paper, we expand the work of~\cite{Petro2023} in three main ways. First,  we adopted a model that describes the jet plasma properties (bulk four-velocity and plasma magnetization) as a function of distance $z$ from the central engine. By accounting for this location-dependent evolution of $\Gamma$ and $\sigma$, we self-consistently link the microphysics of particle acceleration to the macroscopic jet structure. Second, we relax the assumption that magnetic energy dissipation occurs close to the BLR radius, and we account for additional sources of external radiation, namely the accretion disc and dusty torus. Finally, we do not use the empirical relation $\dot{m} \propto \Gamma^{s}$ with $s\sim 3$ as previously done in \cite{2021MNRAS.501.4092R, Petro2023}. 

Nonetheless, the mass accretion rate and the jet bulk Lorentz factor are still related in our model. In ~\cite{Petro2023} $\Gamma$ at the BLR distance was determined by an arbitrary combination of $\mu$ and $\sigma$. For jets with fixed $\mu$, lower values of $\sigma$ at the emitting region translated to higher Lorentz factors $\Gamma$ and higher mass accretion rates $\dot{m}$. The combination of high mass accretion rates (larger energy density of the external photon fields) and high Lorentz factors could further increase the energy density of the external photons in the comoving frame, therefore leading to higher EC emission. In our theoretical framework, the bulk Lorentz factor $\Gamma$ of the emitting region depends on the distance from the SMBH and the initial magnetization $\sigma_0$. 
If one sets the distance of the emitting region to be $z=R_{\rm BLR}$, then this would impose a direct relation between the mass accretion rate and the local Lorentz factor $\Gamma(z)$. Increasing the mass accretion rate would lead to a more extended BLR ($R_{\rm BLR} \propto \dot{m}^{1/2}$) and in turn a larger local Lorentz factor as long as $z_{\rm acc}>R_{\rm BLR}$. This implies that these quantities are not independent in our model. Unlike the previous study, however, we cannot simultaneously fine tune $\sigma$ at the emission site; greater distances for the same $\sigma_{0}$ yield softer particle distributions. 

The results of our extended model can be directly compared to those of~\cite{Petro2023}, revealing how the distance of the dissipation region affects the broadband SED, the neutrino production, and the ratio of the neutrinos with the $\gamma$-rays in the Fermi band. We find that $z/R_{\rm BLR}$ is a crucial parameter governing the SED shape. When the dissipation occurs near the BLR (as in~\cite{Petro2023}), the intense external photon field leads to a pronounced high-energy bump dominated by EC emission. In this regime, BLR photons serve as seed photons for IC scattering. We recover the dichotomy between high-power, Compton-dominated blazars and low-power, neutrino-efficient blazars within a single framework. In ~\cite{Petro2023}, these two classes were treated by choosing different jet parameters at the emitting region (e.g., $\sigma$) without considering an underlying model for their evolution along the jet. Here, by varying the distance $z$ of the emitting region from the SMBH while keeping the initial jet parameters fixed at the jet base, we demonstrate how different manifestations of the SED can be recovered depending on whether the emission occurs close to the accretion disk, at sub-pc, or at pc scales. When $z$ is close to the SMBH (energy density of the magnetic field is dominant) and the magnetization is high $\sigma >4$), the produced spectrum from the emitting region matches an HSP BL Lac object and leads to a neutrino luminosity that is similar and can overshoot the $\gamma$-ray luminosity if $\gamma \gamma$ absorption is included. When $z/R_{\rm BLR}\simeq 1$, the magnetization is lower and the bulk Lorentz factor is close to its asymptotic value, the boosted external field in the comoving system matching an FSRQ object. In this case, the neutrino emission is a fraction of the $\gamma$-ray luminosity in the Fermi band. The resulting neutrino emission peaks at energies of $\sim 1-10$ PeV, placing it within the sensitivity range of current and next-generation neutrino observatories such as IceCube-Gen2 ~\citep{2021JPhG...48f0501A}, and KM3NeT ~\citep{2019APh...111..100A}.

For our baseline model, we find that the CD is always below unity compared to the observational data from \citep{2023ApJ...944..157C}. The synchrotron peak carries most of the energy, even if the emitting region is placed close to the BLR region, where we expect the external photon field to be the dominant source of energy losses for the high-energy particles. In reality, the interactions of the high-energy pairs are happening in the Klein-Nishina, resulting in a less prominent Compton emission in the Fermi band. There are two practical ways for increasing the CD without changing the dissipation distance $z_{\rm acc}$: (i) 

Reduce the jet-to-accretion power ratio $\eta_{\rm j}$. Lower $\eta_{\rm j}$ reduces the magnetic energy density, weakens synchrotron losses, and leaves EC relatively stronger; (ii)  Boost the external photon energy density in the comoving frame. For fixed $\dot m$ and BLR covering factor $\eta_{\rm BLR}$, the only way to do so is to increase the bulk Lorentz factor $\Gamma$, which Doppler-boosts BLR photons in the frame of the emitting region. Achieving a higher $\Gamma$ requires a larger initial magnetization $\sigma_0$, thereby providing more Poynting flux for bulk acceleration.

An important parameter in our model is the particle acceleration efficiency, characterized by $\eta_{\rm acc}$. PIC simulations of relativistic magnetic reconnection typically yield $\eta_{\rm acc}\sim 10$, corresponding to efficient, near-Bohm acceleration with very short acceleration timescales (see Eq.~\ref{Eq:t_acc}) \citep[e.g.,][]{2014ApJ...783L..21S,2015ApJ...806..167G}. Such efficient acceleration naturally predicts synchrotron emission extending up to the synchrotron burnoff limit at $10-100$ MeV due to radiation reaction losses, implying extreme HSP blazars. In contrast, reproducing intermediate- or low-synchrotron-peaked (ISP/LSP) blazars, whose observed synchrotron peaks lie below the keV energy band, requires adopting significantly larger values e.g., $\eta_{\rm acc}\sim10^6$ \citep{2021MNRAS.501.4092R, Petro2023}. Such large values imply a much slower acceleration process, in strong tension with current PIC-based predictions. This discrepancy suggests that additional effects not captured by simplified one-zone models, such as guide-field reconnection, turbulence, or particle escape from reconnection sites, could reduce effective particle acceleration efficiency and thus explain the observed diversity of blazar synchrotron peak energies.
    
Another key consideration is the evolving structure of the jet’s magnetic field (poloidal $B_{\rm p}$ versus toroidal $B_{\phi}$) and its influence on magnetic dissipation. Near the base of the jet the magnetic field is presumed to be dominated by a strong $B_p$ (vertical or guide-field component), with $B_p \gg B_{\phi}$. Magnetic reconnection in this regime would occur with a non-negligible guide field component, which theory and simulations show tends to suppress the acceleration efficiency and reduce the maximum energy of accelerated particles
\citep{2015MNRAS.450..183S,2023ApJ...948...19F,2023ApJ...954L..37L,2024ApJ...964L..21W}. In other words, a strong guide field (as expected near the black hole) can lead to a softer non-thermal particle spectrum and a lower energy cutoff. This could make it difficult for an inner-zone dissipation to produce extremely high-energy photons or neutrinos despite the high magnetization. Further out along the jet, however, the toroidal component $B_{\phi}$ grows in relative importance (due to flux freezing and jet rotation), and at some distance it is expected that $B_{\phi} \gg B_{\rm p}$. In highly magnetized flows this distance is beyond the light surface \citep{Komissarov2007}.  \cite{Vlahakis2004} also showed that in relativistic jets such as 3C~345, the poloidal and toroidal magnetic field components become approximately equal around the Alfvén lever arm radius $\varpi_{\rm A}$, beyond which $B_{\phi}$ dominates. In the $r-$self-similar jet solution, $B_{\phi} = B_{\rm p}$ occurs at each streamline’s Alfvén radius, spanning from $10^{-3}-1$~pc $(10-10^4 R_{\rm s}$) from the SMBH for the 3C~345 parameters. At distances where the toroidal field dominates, magnetic reconnection, if triggered in a knot or shock, can proceed in the weak guide-field limit, efficiently accelerating particles to high energies. Our model qualitatively incorporates this by considering dissipation at various $z$. 

\section{Conclusions}\label{sec:conclusions}

We introduced a comprehensive radiative model for blazar jets, examining photon and neutrino emission powered by magnetic reconnection occurring at varying distances from the central SMBH. By relaxing a fixed dissipation location near the BLR, our model accounts for spatial evolution of jet properties such as magnetization and bulk Lorentz factor, as well as varying external photon fields. We demonstrate that different emission locations along the jet significantly impact the resulting multi-messenger outputs, bridging observational characteristics between low-luminosity BL Lac like sources and high-luminosity, Compton-dominated FSRQs. Specifically, we find that neutrino production is most efficient on sub-parsec scales, where accelerated protons interact with jet-synchrotron and BLR photons. Further out the BLR region, at parsec scales, the reduced photon densities and softer proton distribution lead to inefficient photopion interactions Overall, our results provide a unified framework that can help interpret blazar observations across electromagnetic and neutrino spectra, emphasizing magnetic reconnection as a viable mechanism driving multi-messenger astrophysical phenomena.

\section*{Acknowledgements}
S.I.S. and M.P. acknowledge support from the Hellenic Foundation for Research and Innovation (H.F.R.I.) under the ``2nd call for H.F.R.I. Research Projects to support Faculty members and Researchers'' through the project UNTRAPHOB (Project ID 3013). The authors acknowledge the use of Grammarly to help them improve the writing style and coherence of the text.

\section*{Data Availability}

The inclusion of a Data Availability Statement is a requirement for articles published in MNRAS. Data Availability Statements provide a standardised format for readers to understand the availability of data underlying the research results described in the article. The statement may refer to original data generated in the course of the study or to third-party data analysed in the article. The statement should describe and provide means of access, where possible, by linking to the data or providing the required accession numbers for the relevant databases or DOIs.



\bibliographystyle{mnras}
\bibliography{paper} 




\appendix

\section{Luminosity transferred to the non-thermal charged particles}\label{app_A}
The Poynting luminosity that passes through a spherical surface of radius $r$, where $r \approx R/\sin \theta_{\rm int}$, $R$ is the cross-sectional radius of the jet and $\theta_{\rm int}\equiv\eta_0/\Gamma$, reads $L_{P, \rm iso}=\frac{c}{4\pi} \beta B^2 4 \pi r^2 \approx c \beta B^2 R^2 / \sin^{2}\theta_{\rm int}$. To calculate the electromagnetic power of the jet we multiply $L_{P, \rm iso}$ with $\Delta \Omega/(4\pi)=(1- \cos\theta_{\rm int})/2$, which is the fraction of the sky that the jet subtends,
\begin{equation}
    L_{P,\rm jet} =  L_{P, \rm iso}\frac{\Delta \Omega}{4\pi}=\frac{c \beta \Gamma^2 B'^2 R^2}{2(1+\cos\theta_{\rm int})},
\end{equation} 
where we used the Lorentz transformation for the magnetic field, $B = \Gamma B'$. A fraction $f_{\rm rec}$ of the Poynting jet luminosity $L_{\rm P,jet}(\Omega)=L_{P,\rm jet}/\Delta \Omega$ is transferred to non-thermal particles through magnetic reconnection, 
\begin{equation}\label{App_A}
    L_{\rm i}(\Omega) = f_{\rm rec}L_{\rm P,jet}(\Omega).
\end{equation}
We may connect the comoving injection luminosity, which is the input quantity for the radiative calculations, with the injection luminosity as measured by a stationary observer as follows,
\begin{equation}
    L'_{\rm i}  =  \int d\Omega' L'_{\rm i}(\Omega') = \int d\Omega' \delta^{-4} L_{\rm i}(\Omega) = \int d\Omega \delta^{-2} L_{\rm i}(\Omega).
\label{Eq:Com_L'_i}
\end{equation}  
We then substitute $\delta = \Gamma^{-1} (1-\beta \cos \theta)^{-1}$ and perform the integration in solid angle,
\begin{eqnarray}
L'_{\rm i} & = & \frac{f_{\rm rec} c \beta B'^2 R^2\Gamma^4}{2\sin^{2}\theta_{\rm int}} \int_{\cos\theta_{\rm int}}^1 d\mu (1-\beta \mu)^2  = \nonumber \\ 
& = & \frac{f_{\rm rec} c \beta B'^2 R^2\Gamma^4}{6 \beta \sin^{2} \theta_{\rm int}} \left[\left(1-\beta \cos \theta_{\rm int}\right)^3-(1-\beta)^3\right] 
\label{Eq:Inj_lum}
\end{eqnarray}

For $\theta_{\rm int}=\eta_0/\Gamma \ll 1$, corresponding to small opening angles and high bulk Lorentz factors of the emitting region, the luminosity of the non-thermal particles can be written as follows,

\begin{equation}
    L'_{\rm i}  =  \frac{2}{3}\frac{f_{\rm rec} L_{\rm P,jet}}{8\Gamma^2}\left(\eta_0^4+3\eta_0^2+3\right) \approx \frac{f_{\rm rec} L_{\rm P, jet}}{4\Gamma^2}
\label{Eq:Com_L'_i_approx}
\end{equation}
where we used the approximations $\sin \theta_{\rm int} \approx \theta_{\rm int}$, $\cos \theta_{\rm int} \approx 1-\theta^2_{\rm int}/2$, and $\eta_0 \ll 1$.

\begin{figure}
\centering
\includegraphics[width=0.45\textwidth]{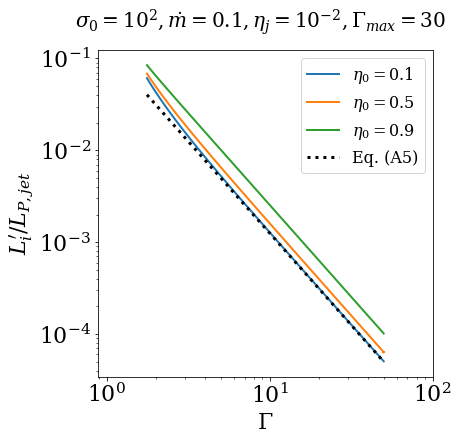} 
\caption{Ratio of comoving injection luminosity of non-thermal particles to the jet Poynting luminosity as a function of the bulk Lorentz factor $\Gamma$ for different values of $\eta_0$. The black dashed line shows the approximate expression of the ratio given by Eq.~\ref{Eq:Com_L'_i_approx}. }
\label{fig:L_i_div_L_P}
\end{figure}

In Fig.~\ref{fig:L_i_div_L_P} we present the ratio of the injected luminosity to the non-thermal particles with respect to the jet Poynting luminosity as a function of the bulk Lorentz factor $\Gamma$ of the emitting region. We observe that the approximate expression works well for small apparent angles ($\eta_0 < 0.5$) and $\Gamma > 3$.

\section{Neutrino production on the BLR photons}\label{app:nu_on_BLR}
Assuming that the target photon field consists of monochromatic BLR photons with comoving energy $E'_{\rm BLR}$ and approximating the cross-section and inelasticity as in \citep{2003ApJ...586...79A}, we find that the photopion efficiency near the threshold becomes,
\begin{equation}
f_{\rm p \pi} \approx \left(1 + \frac{354\,T_{\rm eff}}{\Gamma(z)\,\eta_{\rm BLR}}\right)^{-1}.
\end{equation}
Since $\Gamma(z)$ increases with distance, this implies that, for fixed $T_{\rm eff}$ (assuming that $T_{\rm eff}$ corresponds to the max temperature of the multi-temperature disk)  and $\eta_{\rm BLR}$, the photopion efficiency increases with distance but only while BLR photons remain main photon target for photopion production. Once the emitting region is beyond the BLR, the declining synchrotron photon density becomes the limiting factor. This decrease in target photon density leads to a corresponding drop in neutrino luminosity, explaining the observed suppression of neutrino emission at larger distances from the SMBH in Fig.\ref{fig:baseline_nu_vs_z}.

Efficient photopion interactions are most likely favored when the emitting region is located upstream of the BLR, allowing BLR photons to be Doppler-boosted into the jet’s comoving frame. Besides the usual threshold condition for photopion production, $\gamma'_{\rm p}E'_{\rm BLR}\gtrsim E_{\rm th} \simeq 145$ MeV, where $E'_{\rm BLR}=2.7k_{\rm B}\Gamma(z)T_{\rm eff}$ is the boosted energy of the BLR photons to the jet's comoving frame, an efficient neutrino production requires that most of the proton energy budget is carried by protons that satisfy the threshold. This favours hard proton spectra $p<2$, i.e. local magnetizations $\sigma\gtrsim 3$. The maximum attainable Lorentz factor, $\gamma'_{\rm p,\max}$, follows from the minimum of the acceleration time and the relevant radiative loss and escape timescales (see Sec.~\ref{Subsec:Radiative_losses}). Although the Bethe-Heitler and photopion timescales require knowledge of the photon fields in the emitting region, we find that the dominant timescale between synchrotron and escape for our baseline model is the escape timescale (see Fig.~\ref{fig:min_gamma_vs_z}). By equating the acceleration and the escape timescales, we find that the attained proton Lorentz factor is given by,

\begin{equation}
\begin{split}
\gamma'_{\rm p,sat}(z)&= \frac{2q}{\eta_{\rm acc}\,m_{\rm p}\,c^3}\sqrt{\frac{\eta_{\rm j}\,\dot m\,L_{\rm Edd}}{\eta_{\rm c}\,c}\,\frac{\sigma(z)}{\Gamma(z)\,\beta(z)}}
\\
&\simeq \frac{4\times10^6}{\eta_{\rm acc,4}}\sqrt{\frac{\eta_{\rm j,-1}\,\dot m_{-2}\,L_{\rm Edd,46}}{\eta_{\rm c,-1}}
\,\frac{\sigma(z)}{\Gamma(z)\,\beta(z)}}.
\end{split}
\label{Eq:gamma_p_sat}
\end{equation}

\begin{figure}
\centering
\includegraphics[width=0.5\textwidth]{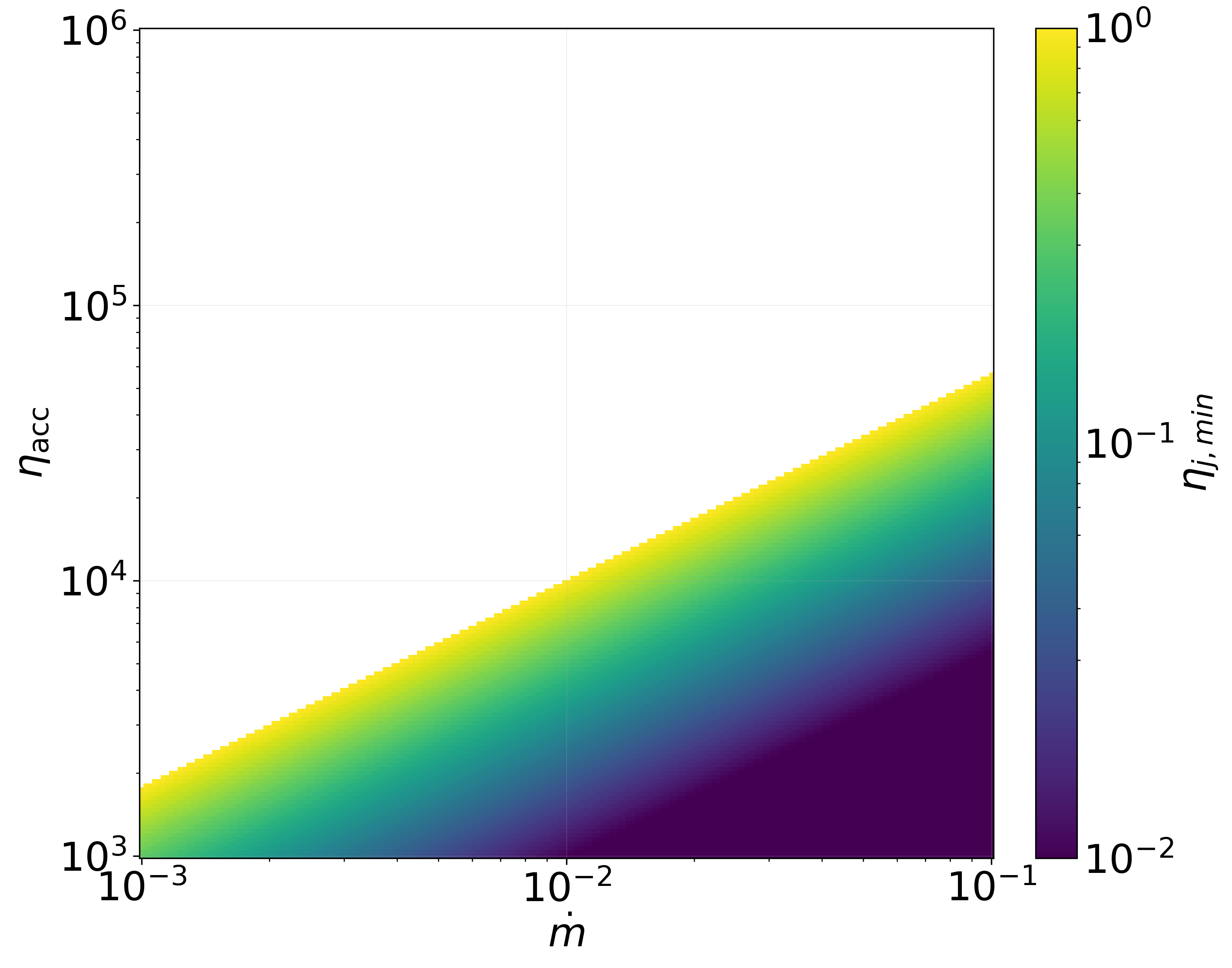} 
\caption{Minimum jet–loading parameter required for photopion interactions between protons of maximum energy and BLR photons. Colour scale: $\eta_{\rm j,\min}(\eta_{\rm acc},\dot m)$ obtained from Eq.~\eqref{eq:nu_thresh} 
for the baseline model. The white area corresponds to $\eta{\rm j}$ that are above unity.}
  \label{fig:eta_thresh}
\end{figure}

Imposing $\gamma'_{\rm p,\max}E'_{\rm BLR}>E_{\rm th}$ and substituting the escape–limited $\gamma'_{\rm p,sat}$ (see Eq. \ref{Eq:gamma_p_sat}) yields the following,

\begin{equation}
\frac{5.4qk_{\rm B}}{\eta_{\rm acc}m_{\rm p}c^{3}}
\left[\frac{\eta_{\rm j}\,\sigma(z)\,\Gamma(z)}{c\,\sigma_{\rm SB}\,\beta(z)}\right]^{1/2}
\!\left[\frac{2\dot m\,m_{p}c^{4}}{\eta_{\rm c}\sigma_{\rm T}}\right]^{3/4}
R_{\rm s}^{1/4}E_{\rm th}^{-1}
\gtrsim 1.
\label{eq:nu_thresh}
\end{equation}
Assuming that $\sigma(z)\Gamma(z)$ is constant which is true as long as $\sigma(z)\gg 1$ and $\beta(z)\sim 1$ in Eq.~\ref{eq:nu_thresh} we can find that the threshold condition is independent of distance but depends only on parameters such as the acceleration efficiency $\eta_{\rm acc}$, the ratio of jet power to accretion power $\eta_{\rm j}$ and the mass accretion rate $\dot{m}$. Figure~\ref{fig:eta_thresh} illustrates the minimum jet-loading parameter ($\eta_{\rm j,\min}$) required to achieve the photopion threshold as a function of these parameters. Increasing the acceleration efficiency $\eta_{\rm acc}$ increases the acceleration timescale and thus lowers the maximum attainable proton energy for a fixed jet power. As a result, achieving efficient proton acceleration at high $\eta_{\rm acc}$ requires stronger magnetic fields, which can be realized through higher jet power ($\eta_{\rm j}$) and mass accretion rates ($\dot{m}$).

\section{External radiation fields}\label{app:external}
External photon fields, which vary with distance $z$ from the SMBH, significantly influence the acceleration of charged particles and the non-thermal radiation emitted by the jet. We take into account the impact of different external photon sources on the jet emission region, considering the contribution of the accretion disc, the Broad Line Region, and a more distant infrared-emitting (IR) dusty torus.
Following \citep{2009MNRAS.397..985G}, we assume that the BLR 
reprocesses and re-emits a fraction of the disc's luminosity. 
We assume a standard Shakura-Sunyaev disc \citep{1973A&A....24..337S} extending from $R_{\rm in}$ to $R_{\rm out}$, with an effective temperature profile,

\begin{equation} 
T_{\rm eff}(R_{\rm d}) = \left( \frac{3 R_{\rm s} \dot{m} L_{\rm Edd}} {16 \pi \eta_{\rm c} \sigma_{\rm SB} R_{\rm d}^3} \left (1-\sqrt{\frac{R_{\rm in}}{R_{\rm d}}}\right) \right)^{1/4}, 
\label{Eq:T_eff}
\end{equation}
where $R_{\rm s}$ is the Schwarzschild radius, defined as $R_{\rm s} = 2 G M_{\rm BH}/c^2$, with $G$ being the gravitational constant and $c$ the speed of light. The parameter $\dot{m}$ represents the dimensionless mass accretion rate, which is the ratio of the actual mass accretion rate to the Eddington accretion rate. The Stefan-Boltzmann is denoted as $\sigma_{\rm SB}$ while $R_{\rm in}$ usually corresponds to the innermost stable circular orbit (ISCO) for a black hole. We fix this distance to $3R_{\rm s}$ and set $R_{\rm out} = 500 R_{\rm in}$. As long as $R_{\rm out}\gg R_{\rm in}$, the actual extent of the disc does not impact our results in any way. 

Each disc annulus emits radiation at a different temperature. A plasma blob in the jet at a distance $z$ moving relative to the accretion disc above the central SMBH receives photons from the various disc annuli at different angles $\theta$. As a result, the Doppler boost of the spectral intensity emitted by its annulus will vary. The specific energy density of disc photons in the comoving frame is,
\begin{eqnarray} 
U^{'}_{\nu', \rm D} & = &  \frac{2 \pi \Gamma}{c} \int_{\mu_{\rm min}}^{\mu_{\rm max}} I_{\nu} \left(1 - \beta \mu \right) {\rm d}\mu \nonumber \\  
& = &  \frac{4 \pi h \Gamma}{c^3}  \int_{\mu_{\rm min}}^{\mu_{\rm max}}  \nu^3 \frac{1 - \beta \mu}{e^{\frac{h \nu}{k_{\rm B} T_{\rm eff}(R_{\rm d})}} - 1} {\rm d}\mu.
\label{Eq:U_disk_com}
\end{eqnarray}
where $\nu$ is the frequency, $\mu = \cos\theta$,  $\theta$ is the angle between the jet axis and each annulus of the accretion disc, and $R_{\rm d}=\sqrt{(1-\mu^2)}z/\mu$ is the disc annulus radius, $\mu_{\rm min,max}=(1+(R_{\rm in,out}/z)^2)^{-1/2}$.

The BLR is modelled as a thin shell that covers the full hemisphere with radius $R_{\rm BLR}=10^{17}L_{\rm d,45}^{1/2}$ cm. We assume that the BLR reprocesses a fraction $f_{\rm BLR}$ of the disc luminosity $L_{\rm d}$, and re-emits it isotropically. Considering the disc as a point source for $R_{\rm BLR} \gg R_{\rm out}$, the intensity of the BLR radiation is then given by \citep{1996MNRAS.280...67G}
\begin{equation}
I_{\rm BLR} = \frac{1}{4\pi} \frac{f_{\rm BLR}L_{\rm d}}{2 \pi R^2_{\rm BLR}} \cos A,
\end{equation}
where $\cos A$ is the inclination angle between the BLR and the AD -- see Eq. 11 in \citep{1996MNRAS.280...67G}. The specific energy density of the BLR in the blob comoving frame is calculated using \citep{1996MNRAS.280...67G}, 
\begin{equation}
U^{'}_{\nu',\rm BLR} = \frac{2 \pi \Gamma}{c} \int_{\mu_{\rm BLR}^{\rm min}}^{\mu_{\rm BLR}^{\rm max}} I^{\rm BLR}_{\nu} \left(1 - \beta_{\Gamma} \mu \right) {\rm d}\mu,
\label{Eq:U_BLR_com}
\end{equation}
where $I^{\rm BLR}_{\nu}$ is the specific intensity from the BLR. The bolometric energy density in the blob rest frame is then given by
\begin{equation}
U^{'}_{\rm BLR} = \frac{2 \pi \Gamma^2}{c} \frac{1}{4\pi} \frac{f_{\rm BLR}L_{\rm d}}{2 \pi R^2_{\rm BLR}}  \int_{\mu_{\rm BLR}^{\rm min}}^{\mu_{\rm BLR}^{\rm max}} \cos A \left(1 - \beta_{\Gamma} \mu \right)^2 {\rm d}\mu,
\label{Eq:U_BLR_com_int}
\end{equation}
The integration limits for $\mu = \cos \theta$ depend on the relative position of the emission region to the BLR,
\begin{equation}
\begin{tabular}{lc} 
$\mu_{\rm BLR}^{\min} = -1, \mu_{\rm BLR}^{\max} = 0$, & $z < R_{\rm BLR}$, \\
$\mu_{\rm BLR}^{\min} = (1 + \eta_{r})^{-1/2}, \mu_{\rm BLR}^{\max} =1$ , & $z > R_{\rm BLR}$,
\end{tabular}
\end{equation}
where $\eta_{r} \equiv z / R_{\rm BLR}$.

The DT is treated as an annular shell of radius $R_{\mathrm{IR}}\simeq 2.5\cdot 10^{18} L^{1/2}_{\rm d,45}cm$ (a few parsec) that re‑emits a fraction $\eta_{\mathrm{IR}}$ of $L_{\mathrm d}$ in the infrared, typically at $T_{\mathrm{IR}}\approx370\,$K. We assume that is a grey-body and following \citet{2009MNRAS.397..985G}, we scale its comoving energy density for distances less than $z<R_{\mathrm{IR}}$,
\begin{equation}
U'_{\mathrm{IR}}(z)=\frac{\eta_{\mathrm{IR}}L_{\mathrm d}}{4\pi R_{\mathrm{IR}}^{2}c}\Gamma^{2}
\label{eq:U_IR}
\end{equation}
while for distances $z>R_{\rm IR}$ we assume that there is no presence of the DT since the bulk Lorentz factor at these distances has reached its asymptotic values and the emission is significantly deboosted.




\bsp	
\label{lastpage}
\end{document}